\documentclass[lettersize,journal]{IEEEtran}
\usepackage{amsmath,amsfonts,amsthm}
\usepackage{algorithm}
\usepackage{array}
\usepackage{textcomp}
\usepackage{stfloats}
\usepackage{url}
\usepackage{verbatim}
\usepackage{graphicx}
\usepackage{cite}
\usepackage{xcolor}
\usepackage{amssymb}
\usepackage{subcaption}
\usepackage[font=small]{caption}
\usepackage{hyperref}
\usepackage{algpseudocode}
\usepackage{siunitx}
\usepackage{balance}
\usepackage{booktabs}

\DeclareSIUnit{\Mbit}{\mega\bit}
\DeclareSIUnit{\Mbps}{\Mbit\per\second}
\DeclareSIUnit{\belmilliwatt}{Bm}
\DeclareSIUnit{\dBm}{\deci\belmilliwatt}

\newcommand{\ind}{1{\hskip -2.5 pt} \mathrm{I}}
\newcommand{\deltaua}{\delta} 

\let\oldnl\nl
\newcommand{\nonl}{\renewcommand{\nl}{\let\nl\oldnl}}
\newcommand{\sinr}{\gamma}
\newcommand{\sinrrep}{\sinr^{\mathrm{rep}}}
\newcommand{\sinrest}{\widetilde{\sinr}}
\newcommand{\sinrestolla}{\widetilde{\sinr}^{\mathrm{OLLA}}}
\newcommand{\mcs}{u}

\newcommand{\bler}{\text{\small\textsc{BLER}}}
\newcommand{\bce}{\text{\small\textsc{BCE}}}
\newcommand{\blertable}{\bler}
\newcommand{\deltaplus}{\Delta^{\textsc{ACK}}}
\newcommand{\deltaminus}{\Delta^{\textsc{NACK}}}
\newcommand{\res}{b}

\newcommand{\nack}{\ind^{\textsc{NACK}}}
\newcommand{\mcsilla}{\mcs^{\mathrm{ILLA}}}
\newcommand{\salad}{SALAD}

\begin{document}
	
\title{SALAD: Self-Adaptive Link Adaptation}

\author{
	\IEEEauthorblockN{
		Reinhard Wiesmayr*,
		Lorenzo Maggi*,
		Sebastian Cammerer,
		Jakob Hoydis,
		Fay\c{c}al A\"it Aoudia,
		Alexander Keller
		\thanks{
			*Equal contribution.\\
			Reinhard Wiesmayr is with ETH Zürich, Gloriastrasse 35, 8092 Z\"urich (Switzerland). His work was conducted during an internship at NVIDIA.\\
			Lorenzo Maggi, Fay\c{c}al A\"it Aoudia, and Jakob Hoydis are with NVIDIA, 10 avenue de l'Arche, 924000 Courbevoie (France).
			Sebastian Cammerer and Alexander Keller are with NVIDIA, Fasanenstraße 81, 10623 Berlin (Germany). Emails: wiesmayr@iis.ee.ethz.ch, \{lmaggi, scammerer, faitaoudia, jhoydis, akeller\}@nvidia.com.\\
			This work has been submitted to the IEEE for possible publication. Copyright may be transferred without notice, after which this version may no longer be accessible.}}

}

\maketitle


\begin{abstract}
	Adapting the modulation and coding scheme (MCS) to the wireless link quality is critical for maximizing spectral efficiency while ensuring reliability. 
	We propose \salad{} (self-adaptive link adaptation), an algorithm that exclusively leverages ACK/NACK feedback to reliably track the evolution of the signal-to-interference-plus-noise ratio (SINR), achieving high spectral efficiency while keeping the long-term block error rate (BLER) near a desired target.
	\salad{} infers the SINR by minimizing the cross-entropy loss between received ACK/NACKs and predicted BLER values.
	Based on this inference, \salad{} selects the MCS via hypothesis testing: if the SINR is likely underestimated, a higher MCS is selected to accelerate link adaptation under improving channel conditions. 
	To prevent BLER drift from its long-term target, \salad{} incorporates a feedback control loop that adjusts the instantaneous BLER target. 
	Over-the-air experiments on a 5G testbed demonstrate that SALAD consistently outperforms the industry-standard outer-loop link adaptation (OLLA). With a single set of parameters, SALAD achieves up to 15\% higher throughput and spectral efficiency than multiple OLLA variants across different traffic regimes, while meeting the BLER target.
\end{abstract}



\section{Introduction}\label{sec:introduction}

Adapting the modulation and coding scheme (MCS) to varying wireless channel conditions, known as link adaptation (LA), is essential for achieving high spectral efficiency (SE). Higher SE translates into increased user throughput, lower latency, and reduced transmission energy thanks to shorter transmission durations and fewer retransmissions.
An effective LA strategy must also guarantee a sufficiently low block error rate (BLER), as excessive re-transmissions waste resources, increase latency, and degrade performance. In fact, transport protocols such as TCP misinterpret packet losses caused by link-level errors as signs of network congestion, resulting in an unwarranted reduction in transmission rate \cite{balakrishnan2002comparison}. 
For these reasons, a target BLER value is typically defined to strike a balance between throughput and reliability. Empirical evidence and theoretical analyses \cite{wu2011coding} support a fixed target BLER value of approximately $10\%$, although adapting the target dynamically based on the current signal-to-interference-plus-noise ratio (SINR) can improve throughput \cite{park2015optimizing}.

LA techniques can be broadly categorized into sampling-based and model-based approaches.
Sampling-based methods, such as auto rate fallback (ARF) \cite{kamerman1997wavelan} and sample rate \cite{bicket2005bit}, adapt the MCS by trial-and-error, solely relying on statistics on transmission successes (ACK) and failures (NACK) for each available configuration, received via HARQ (hybrid automatic repeat request) feedback. 
Due to their slow convergence, these methods are mainly suitable for scenarios with few MCS configurations and relatively static channels, as in Wi-Fi.

In contrast, model-based methods leverage statistical models of the wireless channel and side-information, such as SINR estimates, to make more informed decisions.
A prominent example is outer-loop link adaptation (OLLA) \cite{pedersen2007frequency}, the current industry-standard for cellular networks. 
OLLA selects the MCS that meets a target BLER according to pre-computed BLER tables and the latest SINR estimation. The latter is adjusted by an offset which is updated according to a fixed stepsize rule based on the received ACK/NACKs.

Despite its simplicity, OLLA is grounded in stochastic approximation theory (see Section~\ref{sec:olla}) and can provably drive the long-term BLER toward the desired target.
Although OLLA's fixed stepsize works reasonably well across different mobility scenarios, it excels in none. A small stepsize results in slow adaptation to rapidly changing channel conditions, which is particularly detrimental for short-lived connections \cite{buenestado2014analysis}. 
Conversely, a large stepsize can cause instability in static regimes, leading to SE loss and/or short-term BLER degradation. 
Our contribution aims at combining the best of both worlds: the stepsize for SINR updates increases as the channel changes more rapidly.

SINR estimation is a notoriously difficult challenge, especially for the base station (BS) in the downlink (DL), which is our main focus.
Considering 5G NR with time-division multiplexing (TDD), channel reciprocity allows the BS to estimate the useful signal strength via sounding reference signals (SRS), although interference remains unobservable. An alternative common approach is to let the user equipment (UE) estimate the SINR based on channel state information reference signals (CSI-RS) and report it to the BS using the channel quality indicator (CQI). Conversely, in frequency-division multiplexing (FDD), the only option is to rely on CQI reports \cite{3gpp.38.214}.
Yet, CQI reports present several well-known limitations. First, their granularity is coarse, as the CQI is an integer ranging from \num{0} to \num{15}. Second, the reports can be infrequent, with periodicity configurable from a few to several hundreds of slots. Moreover, even when frequent, CQI reports are still delayed in TDD systems, as they can only be transmitted during uplink slots. Finally, reports are wide-band, while transmission often occurs over a narrow band.
Motivated by this, we reduce the reliance on CQI reports by introducing a principled method to infer the SINR using only ACK/NACK feedback. \\

\subsection{Problem formulation}

To formalize the link adaptation problem, we consider a single user and denote its effective\footnote{The \emph{effective} SINR, or simply called SINR throughout the text, of a user scheduled on different time-frequency resources is defined as the SNR of an equivalent AWGN channel that yields the same BLER as the original channel.} signal-to-interference-plus-noise ratio (SINR) by $\sinr_t$. We consider that the transport block size (TBS) $\res_t$ is provided by the MAC scheduler.\footnote{In practice, the MCS selection may occur prior to TBS allocation. In this case, the TBS can be estimated from past allocations.}
At each slot $t$, the MCS index $\mcs_t$ is ideally selected to maximize the expected SE, defined as the SE\footnote{The SE is defined as the modulation order, i.e., the number of bits per symbol, multiplied by the code rate.} of the selected MCS $\mcs$ multiplied by the probability of successful transmission, while not exceeding the predefined BLER target $\tau\in(0, 1)$:
\begin{align}
	& \max_{\mcs \in U} \, \mathrm{SE}(\mcs) \times \left( 1 - \bler(\mcs,\sinr_t,\res_t) \right) \label{eq:pb} \\
	& \ \mathrm{s.t.} \ \bler(\mcs,\sinr_t,\res_t) \le \tau, \notag 
\end{align}
where $\mathrm{SE}(\mcs)$ is the SE corresponding to MCS $\mcs$ and $U$ is the discrete set of available MCS indices.

If the SINR $\sinr_t$ is known perfectly, then \eqref{eq:pb} can be solved by brute force search. Indeed, the MCS's SE is known, while the BLER can be reliably pre-computed via simulations on an additive white Gaussian noise (AWGN) channel, for various combinations of SE, SINR, and TBS.
In practice, however, the true SINR is unknown and can only be estimated, with accuracy varying across scenarios, as discussed in Section~\ref{sec:introduction}. 
As a result, link adaptation requires a non-trivial joint inference of the unknown SINR and control of the selected MCS at each slot.

\subsection{Our contribution}

We propose self-adaptive link adaptation (\salad), a novel algorithm that estimates the SINR by minimizing the cross-entropy loss between the observed ACK/NACK feedback and the BLER via gradient descent, as described in Section~\ref{sec:inference}. 
When the SINR is deemed to be underestimated---based on a hypothesis testing---\salad{} increases the \emph{instantaneous} BLER target to probe a higher MCS value and corrects the SINR estimation accordingly (Section~\ref{sec:mcs_selection}). 
To meet a desired \emph{long-term} BLER, \salad{} adjusts the instantaneous BLER target via a feedback loop (Section~\ref{sec:bler_feedbackloop}). 

We demonstrate that \salad{} outperforms OLLA in system-level simulations and over-the-air experiments conducted within a controlled setup comprising a rotational absorber table between a commercial off-the-shelf (COTS) open-RAN radio unit (O-RU) and a COTS user equipment (UE) in an anechoic chamber. 
In Section~\ref{sec:results}, we show that, with a single set of pre-defined parameters, \salad{} achieves up to \num{15}\% higher throughput and SE compared to multiple variants of OLLA with different stepsizes, all while reliably meeting the BLER target. Moreover, \salad{} is more robust to delayed CQI reports than OLLA. 
Finally, parameter tuning allows SALAD to adapt to extreme throughput scenarios, yielding up to an additional \num{3}\% throughput gain.

An implementation of SALAD is publicly available at \cite{ourcode}.


\begin{table}
	\centering
	\caption{List of symbols.}
	\label{tab:example}
	\begin{tabular}{@{}l l@{}}
		\toprule[1pt]
		\textbf{Symbol} & \textbf{Meaning} \\
		\midrule[.5pt]
		$t$ & Upcoming slot index \\
		$\sinr$ & True signal-to-interference-plus-noise ratio (SINR) \\
		$\sinrrep$ & SINR corresponding to the last CQI report \\
		$\blertable$ & Block error rate (BLER) from pre-computed table \\
		$\tau$ & Long-term BLER target \\
		$\nack$ & NACK indicator (1 if NACK, 0 if ACK) \\ 
		$\mcs$ & Modulation and coding scheme (MCS) index \\
		$\mathrm{SE}(u)$ & Spectral efficiency of MCS $\mcs$ \\
		$\res$ & Transport block size (TBS) \\
		$\deltaua$ & Number of in-flight packets \\
		$\sigma(.)$ & Logistic sigmoid function \\
		$c,s$ & Sigmoid center/scale for BLER approximation \\
		$\sinrestolla$ & OLLA's estimated SINR \\
		$\Delta$ & OLLA's SINR offset \\
		$\deltaminus$ & OLLA's offset stepsize adjustment upon a NACK \\
		$\deltaplus$ & OLLA's offset stepsize adjustment upon an ACK \\
		$\sinrest$ & \salad's estimated SINR \\
		$\varepsilon$ & \salad's learning rate \\
		
		$\theta$ & \salad's teacher model parameters ($\theta\in \mathbb{R}^K$) \\
		
		$\beta$ & \salad's teacher model regularization coefficient \\
		
		$E$ & \salad's integral BLER error \\
		$k_E$ & \salad's integral BLER error coefficient \\
		$\tau_t$ & \salad's instantaneous BLER target at slot $t$ \\ 
		$\tau^{\mathrm{probe}}$ & \salad's high BLER target for MCS probing \\
		$\mathcal H_0$ & Null hypothesis that SINR estimates are correct \\
		$\mathcal S$ & \salad's bias score \\
		$\rho$ & \salad's bias score ratio threshold for MCS probing \\
		$T$ & \salad's bias score window length (n. samples) \\
		$p^{\mathrm{probe}}$ & \salad's probing probability if score ratio $>\rho$ \\
		
		\bottomrule[1pt]
	\end{tabular}
\end{table}

\subsection{Related work}

In the context of cellular networks, OLLA was first proposed for link adaptation in the early 2000s \cite{nakamura2002adaptive, pedersen2007frequency}, although its mechanism and theoretical justification via Markov chain arguments dates back to the 1990s, initially applied to power control \cite{sampath1997setting}.
Several heuristic extensions of OLLA have since been proposed. These include stepsize adaptation \cite{delgado2017fast, zhu2023nolla}, offset dynamics optimization via reinforcement learning (RL) \cite{kela2022reinforcement}, initialization strategies of the offset based on prior connection statistics \cite{duran2015self}, parameter tuning through deep learning \cite{mandelli2021troll}, enhanced feedback mechanisms for ultra-reliable connections \cite{peralta2022outer}, and customization for emerging applications such as extended reality \cite{paymard2022enhanced}. 
Alternative link adaptation strategies with firm theoretical foundations formulate the problem as a multi-armed bandit (MAB), where each MCS corresponds to an ``arm'' \cite{combes2014optimal, pulliyakode2017reinforcement}. However, these methods often exhibit slow convergence, similar to other sampling-based methods originally designed for Wi-Fi, such as ARF \cite{kamerman1997wavelan} and sample rate \cite{bicket2005bit}. 
To address this limitation, \cite{saxena2019contextual} formulates a contextual MAB framework that jointly learns the BLER for each of the MCS configuration given the current network context.
In \cite{saxena2020bayesian}, a Bayesian approach is used to learn the BLER for each MCS, with Thompson sampling guiding the MCS selection process. 
Subsequently, \cite{saxena2021reinforcement} proposes latent Thompson sampling to infer the underlying SINR. However, such approaches are not designed to accommodate rapid SINR variations and do not explicitly target a predefined BLER. 
A supervised learning framework is introduced in \cite{martin2021emerging} to directly predict the BLER.
A practical approach that combines deep learning for BLER prediction and a calibration method to compensate for prediction errors is studied in \cite{huang2021deluxe}. The work in \cite{peri2024offline} employs offline RL approach via decision transformers under the assumption of persistent CQI reporting.


%


\section{The industry standard: OLLA} \label{sec:olla}

Since the adoption of 3G, the industry standard for link adaptation in cellular systems is the outer-loop link adaptation (OLLA) algorithm \cite{pedersen2007frequency}. 
To drive the long-term BLER close to the target value $\tau$, OLLA maintains an additive offset variable $\Delta$ [\SI{}{\dB}] that corrects the latest SINR estimate $\sinrrep$. The latter can be obtained by CQI reports as explained in \cite{3gpp.38.214}. 
When a NACK is received via the HARQ feedback (denoted by the indicator function $\nack=1$), the offset is decreased by a predefined value $\deltaminus$. Conversely, an ACK ($\nack=0$) leads to a offset increase of $\deltaplus$, computed as:
\begin{equation}
	\deltaplus = \frac{\tau}{1-\tau} \deltaminus.
\end{equation}
We now fix slot $t$. Let $\deltaua_t$ be the number of in-flight packets at slot $t$, for which the HARQ feedback has not been received yet. Then, the latest SINR estimate available at slot $t$ is $\sinrest_{t-\deltaua_t}$, computed as in line 12 of Algorithm \ref{alg:olla}.
Then, OLLA assigns to the user the largest MCS whose BLER does not exceed the target $\tau$, according to the latest SINR estimate:
\begin{align} 
	\mcsilla(& \sinrestolla_{t-\deltaua_t},\tau, \res_t) := \notag \\
	& \operatorname{argmax} \big\{ \mcs \!: \blertable\!\left(\mcs, \sinrestolla_{t-\deltaua_t}, \res_t\right) \le \tau \big\}, \label{eq:ILLA}
\end{align}
where $\blertable$ contains pre-computed BLER values, simulated on the AWGN channel. This step is commonly called inner-loop link adaptation (ILLA)\footnote{An alternative, drop-in replacement for ILLA selects the MCS that maximizes the expected spectral efficiency in~\eqref{eq:ILLA}.}.

\begin{algorithm}
	\caption{Outer-Loop Link Adaptation (OLLA), e.g., \cite{pedersen2007frequency}} \label{alg:olla}
	\begin{algorithmic}[1]
		\State \textbf{Parameters} BLER target $\tau$, NACK stepsize $\deltaminus$.
		\State Set ACK offset update $\deltaplus=\tfrac{\tau}{1-\tau} \deltaminus$
		\State Initialize $\Delta := 0$, $\sinrrep := 0$
		\For{slot $t=0,1,\dots$}
		\State Receive set $\mathcal N$ of ACK/NACK and CQI report (if any). 
		
		\If{CQI is reported}
			\State Convert it to SINR $\sinrrep$ [\SI{}{\dB}]
		\EndIf

		\For{$\nack \in \mathcal{N}$}
		\State Update the SINR offset as:
		\begin{equation} \label{eq:olla_update}
			\Delta \leftarrow \left\{
			\begin{array}{ll}
				\!\!\! \Delta + \deltaplus & \text{if } \nack =0 \ \ \text{(ACK)} \\
				\!\!\! \Delta - \deltaminus & \text{if } \nack =1 \ \ \text{(NACK)}
			\end{array}
			\right.
		\end{equation}
		\EndFor
	\State Set the latest SINR estimate to $\sinrestolla_{t-\deltaua_t}\!:=\!\sinrrep \!+\! \Delta$ [\SI{}{\dB}]
	\If{user is scheduled}
		\State Select the MCS as $\mcsilla(\sinrestolla_{t-\deltaua_t},\tau,\res_t)$
	\EndIf
	
	\EndFor
	\end{algorithmic}
\end{algorithm}

\subsection{OLLA through the lens of stochastic approximation}
 
To build a solid foundation for our algorithm, it is instructive to reinterpret OLLA within the context of stochastic approximation (SA). 
Consider an unknown non-decreasing function $f$ and a constant target $\tau$, and assume that the equation $f(y)=\tau$ has a root at $y=y^*$ that we want to find. 
The function $f$ can only be observed via noisy samples $\widetilde{f}$, providing an unbiased estimate of the true value, i.e., $\mathbb{E}[\widetilde{f}(y)]=f(y)$ for all $y$. 

SA starts with an initial guess $y_0$ of the root $y^*$; at time $t> 0$, upon observing $\widetilde{f}(y_{t-1})$, it updates its guess as \cite{robbins1951stochastic}:
\begin{equation} \label{eq:sa}
	y_{t} = y_{t-1} + \alpha_{t} \left( \tau - \widetilde{f}(y_{t-1}) \right),
\end{equation}
where $\alpha_t>0$ is SA's stepsize. If $\widetilde{f}(y_{t-1})$ undershoots the target $\tau$, i.e., $\widetilde{f}(y_{t-1})<\tau$, then $y_{t-1}$ is likely an underestimation of the root $y^*$, since $f$ is non-decreasing. Thus, $y_{t}>y_{t-1}$, and their difference $y_{t}-y_{t-1}$ is proportional to the undershoot. The reverse argument holds in case of an overshoot. 

OLLA is indeed an SA algorithm under two assumptions: 
\begin{itemize}
	\item[i)] the user's SINR report is fixed, e.g., $\sinrrep:=0$ 
	\item [ii)] the ACK/NACK feedback $\nack_t$ is received immediately upon MCS selection $\mcs_t$, i.e., $\delta_t=0$ for all $t$.
\end{itemize}
In this case, the SINR estimate $\sinrestolla_t$ plays the role of $y_t$, and the unknown function $f$ at time $t$ is
\begin{equation}
	f_t(\sinrestolla_t):= \bler \left(\mcsilla(\sinrestolla_t, \tau,\res_t), \sinr_t, \res_t \right),
\end{equation}
which is time-varying due to the true SINR $\sinr_t$ and the TBS $\res_t$.
The function $f_t$ is observed via the feedback $\nack_t$, which plays the role of the noisy observation $\widetilde{f}_t(\sinrestolla_t)$. The latter is indeed unbiased since $\mathbb{E}[\nack_t]=f_t(\sinrestolla_t)$.

The parallel between SA and OLLA becomes clear once we rewrite the offset update \eqref{eq:olla_update} as:
\begin{equation} \label{eq:olla_update_sa}
	\sinrestolla_t = \sinrestolla_{t-1} + \frac{\deltaminus}{1-\tau} \left( \tau - \nack_{t-1} \right) \quad \forall\, t\ge 0.
\end{equation}
In fact, \eqref{eq:olla_update_sa} and \eqref{eq:sa} show the same structure, with the term $\frac{\deltaminus}{1-\tau}$ in \eqref{eq:olla_update_sa} acting as the SA's \textit{constant} stepsize $\alpha$.

This parallel confirms that OLLA is effectively designed to achieve a predefined BLER target $\tau$. We proceed by discussing OLLA's stepsize choice under the perspective of SA.

\begin{figure}
	\centering
	\includegraphics[width=\linewidth]{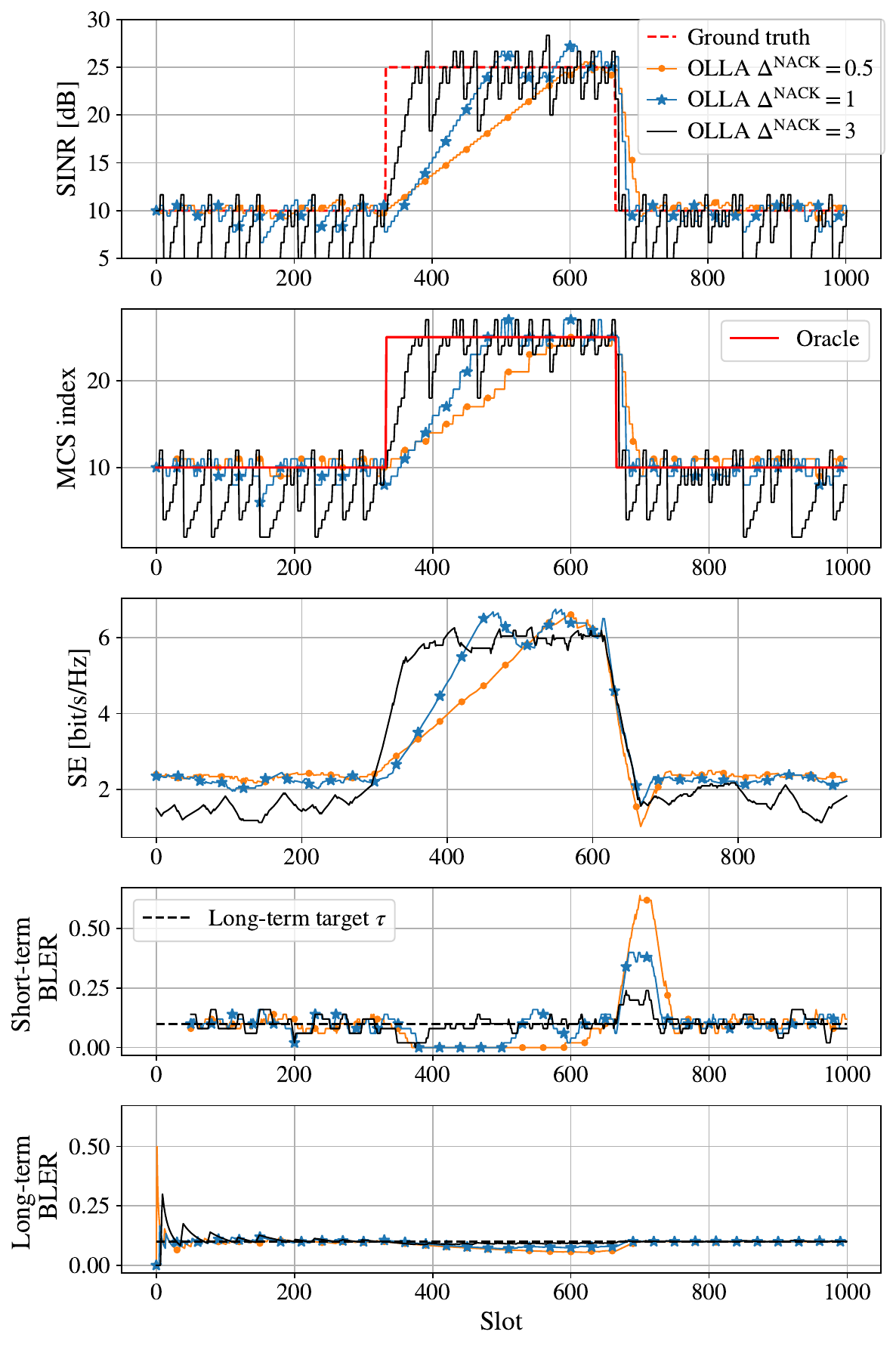}
	\caption{\textbf{OLLA's behavior} when the SINR switches between two values and HARQ feedback delay $\delta=5$. For small values of $\deltaminus$, the MCS adapts slowly to the channel conditions, and the BLER (over a sliding window of 50 slots) exhibits high variance. For large $\deltaminus$, OLLA tracks the SINR variations more rapidly but introduces excessive MCS fluctuations in the stationary phase, causing SE degradation. The oracle knows the ground truth. The code used to produce this result is available at \cite{ourcode}.}
	\label{fig:olla3stepsizes}
\end{figure}

\subsection{Fixed stepsize: Strengths and weaknesses}

If the unknown function $f$ is time-invariant and the stepsize $\alpha_t$ decreases to 0 sufficiently slowly over time (specifically, $\sum_t \alpha_t = \infty$ and $\sum_t \alpha_t^2=0$), then the SA iterates $y_t$ will converge to the root $y^*$ almost surely \cite{blum1954approximation}.

Conversely, if the step size is constant, i.e., $\alpha_t=\alpha$ for all $t$, and $f$ remains time-invariant, the sequence of iterates $y_t$ will form a Markov chain which, under suitable conditions \cite{dieuleveut2020bridging}, will converge exponentially fast to its stationary distribution, with a rate of convergence proportional to the step size $\alpha$. 
Hence, in this \textit{transient} phase, higher values of $\alpha$ result in faster convergence. 
Once the stationary distribution is reached, the Markov chain fluctuates around the root $y^*$ within a region of size $\mathcal O(\alpha)$. Thus, lower values of $\alpha$ help reducing the iterates' variance in the \textit{stationary} phase.

When the unknown function $f$ is time-dependent, using a constant stepsize is a common design choice to effectively track the time-varying root, as in the case of OLLA.
Yet, the considerations above also suggest that there is no universal stepsize value that performs well for OLLA in all scenarios:
\begin{itemize}
	\item When the SINR changes abruptly, a large stepsize is needed to follow the rapid root fluctuations;
	\item When the SINR stabilizes, a smaller stepsize is preferable to avoid significant MCS fluctuations, which would degrade the achieved spectral efficiency. Indeed, when OLLA overshoots by selecting a high MCS, the offset is harshly penalized, causing the MCS to drop well below the optimal value. This behavior is depicted in Figure \ref{fig:olla3stepsizes}.
\end{itemize}

Our proposed \salad{} algorithm is designed to combine the advantages of both strategies: it increases the speed of link adaptation during abrupt SINR changes, and reduces it when the SINR stabilizes, thereby ensuring accurate channel inference while mitigating inefficient MCS fluctuations.

\subsection{The impact of delayed HARQ feedback}

The analogy between OLLA and SA holds in the absence of HARQ feedback delay, i.e., $\deltaua=0$. Yet, in real systems, feedback delay is unavoidable and can lead to performance degradation. 
For instance, suppose that the channel quality deteriorates at time $t$, but the corresponding ACK/NACKs are reported at $t+\delta_t$. 
During the interval $[t,t+\delta_t]$, OLLA continues to select a high MCS, resulting in a series of NACKs received after $t+\delta_t$. This causes the SINR estimate to drop sharply, potentially well below the actual SINR value, ultimately causing the selection of over-conservative MCS.

\salad{} is more robust to HARQ feedback delay, because it updates the SINR estimate using the likelihood of receiving an ACK/NACK, instead of applying a fixed additive step.

%
%


\section{\salad: Self-Adaptive Link Adaptation} \label{sec:our_algo}

Our link adaptation algorithm, called \salad \ (self-adaptive link adaptation), consists of three modules: 
\begin{itemize}
	\item \textbf{SINR inference} (Section~\ref{sec:inference}): The SINR is estimated by binary cross-entropy loss minimization. Unlike OLLA, the SINR update is adaptive and depends on the discrepancy between the ACK/NACK observations and their expected likelihood.
	\item \textbf{MCS selection} (Section~\ref{sec:mcs_selection}): When the SINR is deemed to be underestimated---via hypothesis testing---a high MCS value is probed. This creates a virtuous cycle, enabling faster tracking of sudden SINR surges.
	\item \textbf{Long-term BLER target enforcement} (Section~\ref{sec:bler_feedbackloop}): A feedback-loop mechanism enforces the BLER target $\tau$, compensating for excessive probing when necessary.
\end{itemize}

\begin{figure}
	\centering
	\includegraphics[width=\linewidth]{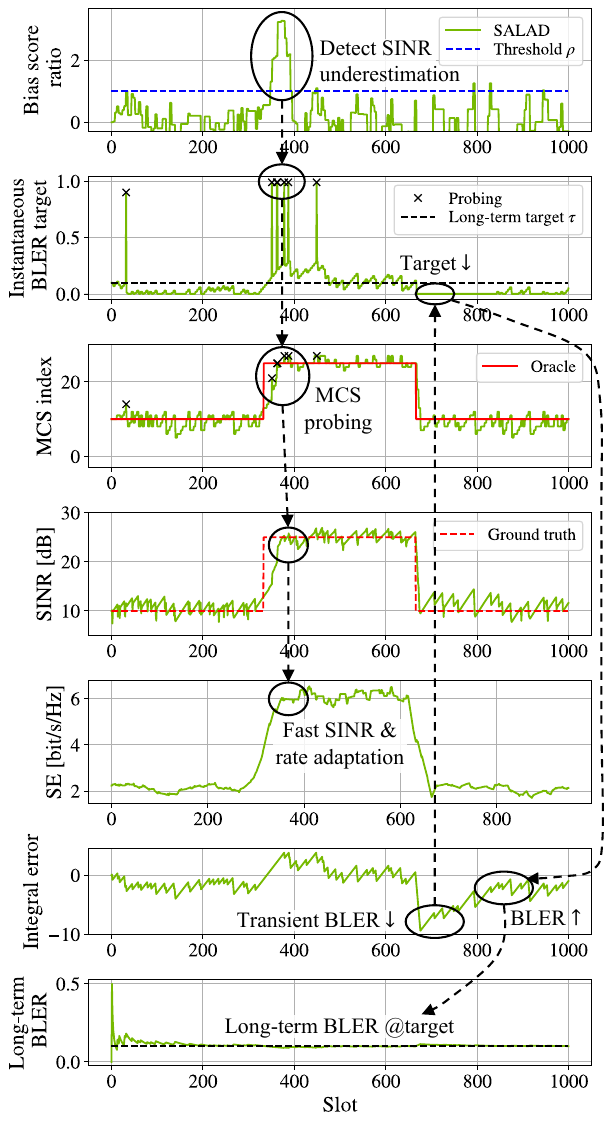}
	\caption{\textbf{\salad's main features}. High values of the bias score ratio indicate SINR underestimation, triggering the probing of high instantaneous BLER target $\tau_t$ and corresponding MCS indices. This results in a quick rate and SINR adaptation. The instantaneous BLER target $\tau_t$ is adjusted according to the integral error $E_t$, steering the long-term BLER towards its target $\tau$.}
	\label{fig:salad_intro}
\end{figure}

\begin{figure*}
	\centering
	\includegraphics[width=\linewidth]{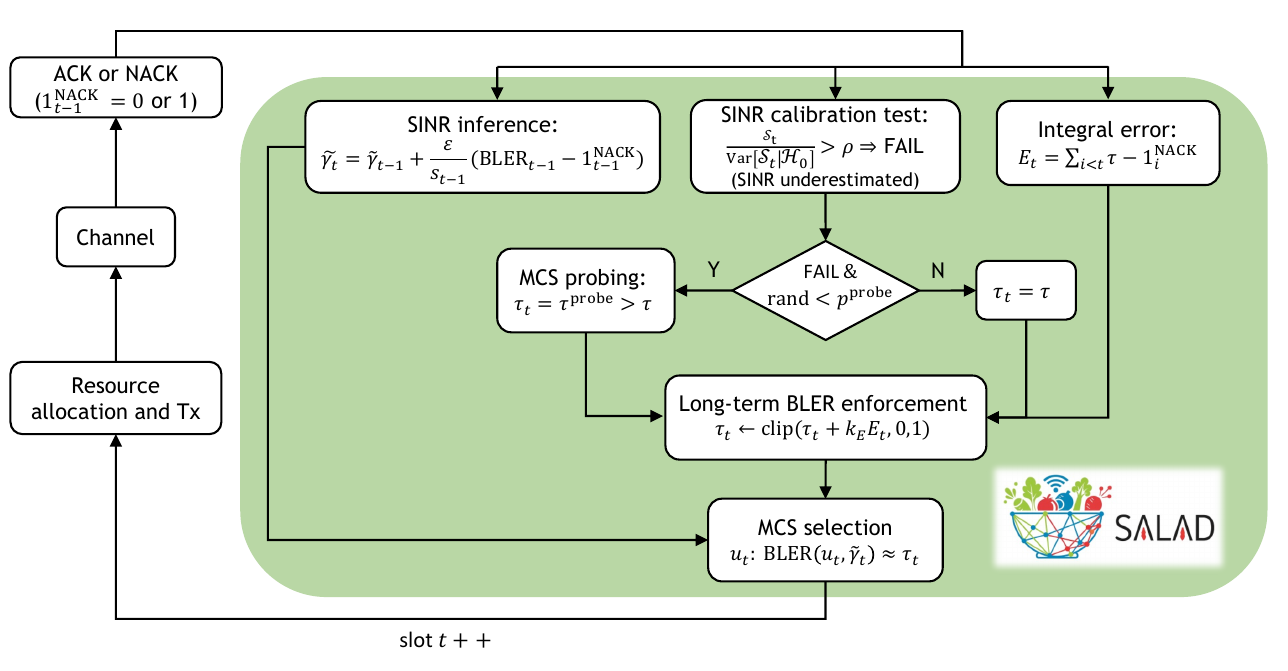}
	\caption{\textbf{\salad's workflow diagram.}}
	\label{fig:salad_diagram}
\end{figure*}

A schematic diagram of \salad{} is provided in Figure~\ref{fig:salad_diagram}, its pseudo-code is shown in Algorithm \ref{alg:salad}, while \salad's qualitative behavior is illustrated in Figure~\ref{fig:salad_intro}.

\subsection{SINR inference} \label{sec:inference}

We illustrate how \salad{} produces an estimate $\sinrest_t$ of the SINR at slot $t$ from past MCS values and the corresponding ACK/NACK feedback. 

First, we quantify the discrepancy between the observed ACK/NACKs in slot $i$ and the SINR estimate $\sinrest_i$ via the empirical binary cross-entropy (BCE) loss, defined as
\begin{align}
	\bce_i(\sinrest_i) := & \, - \nack_i \log \bler\left( \mcs_i, \sinrest_i, \res_i \right) \notag \\
	- \big( 1- & \nack_i \big) \log \bler \left( \mcs_i, \sinrest_i, \res_i \right).  \label{eq:bce}
\end{align}
We seek to minimize the sum of the BCE loss terms across slots:
\begin{equation} \label{eq:opt_pb}
	\min_{\{\sinrest\}_{i\ge 1}} \ \sum_{i\ge 1} \bce_i(\sinrest_i).
\end{equation}
We need to solve \eqref{eq:opt_pb} in an online setting: to estimate the SINR for the next slot, we can only rely on past ACK/NACK feedback.
To this aim, we shall resort to the framework of online optimization. 
To obtain a closed-form expression of the SINR update, we conveniently approximate $1-\bler$ as a suitably translated and scaled sigmoid $\sigma$, as in \cite{carreras2018link}:
\begin{align}
	1\!-\!\blertable(\mcs, \sinr,\res) \!\approx \!\left( 1 + e^{-\frac{\sinr - c(\mcs,\res)}{s(\mcs,\res)}} \right)^{-1} \!\! := \! \sigma_{\mcs,\res}(\sinr), \label{eq:sigmoid_approx}
\end{align}
where $c(\mcs,\res)$ and $s(\mcs,\res)$ are the center and scale of the sigmoid, respectively, which can be obtained via simple mean-square curve fitting. When applied on the BLER tables provided by Sionna \cite{sionna}, such approximation results in an average mean square loss of \num{4e-4}, which is acceptable since our BLER target value $\tau$ is on the order of \num{e-1}. Table \ref{tab:sigmoid_params} reports a few cases, while the full set of results is available at \cite{ourcode}.

\begin{table}[h]
	\caption{Example center and scale parameters of the sigmoid approximation \eqref{eq:sigmoid_approx} for the BLER for different code block size (CBS), considering PDSCH and MCS table index 2, see Section 5 in \cite{3gpp.38.214}.}
	\label{tab:sigmoid_params}
	
	\centering
	\begin{tabular}{@{}l l l l@{}}
		\toprule[1pt]
		\textbf{MCS index} $\mcs$ & $\textbf{CBS}$ (n. bits) & \textbf{center} $c$ & \textbf{scale} $s$ \\
		\midrule[0.5pt]
		2 & 100 & -1.91 &	0.44 \\
		\hline
		2 & 2000 & -2.01  &  0.36  \\ 
		\hline
		6 & 100 & 4.84	& 0.51  \\
		\hline
		6 & 2000 &  5.04  &  0.20  \\
		\hline
		10 & 100 & 8.36 & 0.52  \\
		\hline
		10 & 2000 &  9.04  & 0.04   \\
		\hline
		14 & 100 & 12.20 & 0.57 \\
		\hline
		14 & 2000 & 12.32 & 0.38  \\
		\hline
		20 & 100 & 18.10 & 0.69 \\
		\hline
		20 & 2000 & 18.54 &	0.06  \\
		\bottomrule[1pt]
	\end{tabular}
	
\end{table}

Under approximation \eqref{eq:sigmoid_approx}, the BCE loss is convex. We can then apply the online convex optimization paradigm \cite{hazan2016introduction}.
Specifically, we infer the SINR at slot $t$ via online gradient descent as:
\begin{equation} \label{eq:sinr_update1}
	\sinrest_{t}=\sinrest_{t-1} - \varepsilon \frac{\partial}{\partial\,\sinrest_{t-1}} \bce_{t-1}(\sinrest_{t-1}) \quad \forall\, t > 0
\end{equation} 
where $\varepsilon>0$ is the learning rate and $\sinrest_0$ is initialized arbitrarily. 
We can further express the loss gradient analytically and rewrite \eqref{eq:sinr_update1} in closed-form:
\begin{equation} \label{eq:salad_sinr_update}
		\sinrest_{t}=\sinrest_{t-1} + \varepsilon s_{t-1}^{-1} \left( \blertable_{t-1} - \nack_{t-1} \right) \quad \forall \, t > 0,
\end{equation}
where $\blertable_{t} := \blertable \left(\mcs_{t},\sinrest_{t}, \res_{t}\right)$ and $s_t:=s(\mcs_t,\res_t)$.\\

\noindent \emph{Surprise effect:} We observe that the SINR increment in \eqref{eq:salad_sinr_update} is proportional to the term $\blertable_{t-1} - \nack_{t-1}$ which quantifies the ``surprise'' of observing $\nack_{t-1}$ relative to its expected likelihood $\blertable_{t-1}$. Indeed, 
\begin{itemize}
	\item Suppose $\blertable_{t-1}\approx 1$:
	\begin{itemize}
		\item If an ACK is unexpectedly received ($\nack_{t-1}=0$), then $\sinrest$ is increased significantly, confirming that the SINR is currently underestimated.
		\item Receiving a NACK aligns with expectations and has virtually no impact on the SINR update;
	\end{itemize}
	\item Conversely, if $\blertable_{t-1}\approx 0$:
	\begin{itemize}
		\item A NACK indicates SINR overestimation, resulting in a significant SINR decrease;
		\item an ACK only slightly affects the SINR estimation.
	\end{itemize}
\end{itemize}

\noindent \emph{Remark:} \salad's SINR inference is agnostic to CQI reports. Yet, one can incorporate them as in OLLA, by interpreting $\sinrest$ as an offset to be applied to the most recent CQI report.\\



\subsection{MCS selection} \label{sec:mcs_selection}

\salad{} selects the MCS $\mcs_t$ at slot $t$ via ILLA under an optimized \emph{instantaneous} BLER target $\tau_t$, i.e., 
\begin{equation}
	\mcs_t:=\mcsilla(\sinrest_{t-\deltaua_t}, \tau_t, \res_t) \quad \forall\, t\ge 0.
\end{equation}
Our dynamic target selection philosophy is the following: 
\begin{itemize}
	\item If the SINR inference is considered accurate at time $t$, we select the instantaneous target optimally as $\tau_t=\tau$.
	
	\item If the SINR is detected to be underestimated, a high instantaneous BLER target $\tau_t:=\tau^{\mathrm{probe}} \gg \tau$ is probed. This corresponds to a high MCS index and a high BLER ($\blertable_t\approx 1$). Receiving an ACK ($\nack_t=0$) in this setting provides two advantages: 
	\begin{itemize}
		\item[i)] The spectral efficiency immediately increases;
		\item[ii)] Thanks to the ``surprise effect'', the inferred SINR quickly catches up with the ground truth. This creates a virtuous circle that improves the accuracy of future MCS selections.
	\end{itemize}
\end{itemize}

We now turn to the design of a method for detecting SINR underestimation. \\

\noindent \textbf{Detection of SINR underestimation and MCS probing}. The quality of SINR estimation can be assessed by comparing the BLER of recent transmissions (estimated from pre-computed tables under the assumption of accurate SINR) and the corresponding ACK/NACKs, providing an empirical BLER estimate.
Intuitively, if the BLER significantly exceeds the observed portion of NACKs, the former is being overestimated, while the SINR is underestimated.
To formalize this, we use a simple statistical hypothesis test. 

We first introduce the \emph{calibration bias score}, summing up all signed differences between the BLER values and the ACK/NACK observations $\nack$ over the last $T$ samples:
\begin{equation} \label{eq:score}
	\mathcal{S}_t = \frac{1}{T} \sum_{i= 1}^{T} \blertable_{t-\deltaua_t-i} - \nack_{t-\deltaua_t-i}.
\end{equation}
The score has zero mean under the null hypothesis $\mathcal H_{0,t}$ that all BLER and SINR estimates are correct, where
\begin{equation}
	\mathcal H_{0,t}:=\!\left\{\mathbb E [\nack_{i-\deltaua_t}]\!=\!\blertable_{i-\deltaua_t}, \  t-T\le i \le t-1 \right\}.
\end{equation}
We reject the null hypothesis $\mathcal H_0$, and conclude that the SINR estimation is poor, when the value of $\mathcal S_t$ is unlikely under $\mathcal H_0$. More specifically, we deem the SINR underestimated when $\mathcal{S}_t$, normalized by its variance under $\mathcal H_0$, exceeds a design parameter $\rho>0$:
\begin{align}
	\frac{\mathcal{S}_t}{\sqrt{\mathrm{Var}[\mathcal{S}_t | \mathcal H_{0,t}]}} > \rho \ \ & \ \ \Rightarrow \ \ \text{SINR is underestimated,}  \label{eq:probe_condition} 
\end{align} 
where
\begin{equation} \label{eq:var_St}
	\mathrm{Var}\left[\mathcal{S}_t | \mathcal H_{0,t} \right] = \frac{1}{T^2} \sum_{i=1}^{T} \blertable_{t-\deltaua_t-i} (1 - \blertable_{t-\deltaua_t-i}).
\end{equation}
In this case, as motivated above, a high BLER target $\tau_t:=\tau^{\mathrm{probe}} \gg \tau$ is selected, resulting in the probing of an aggressive MCS value.\\

\noindent \emph{Remark}. A non-negligible HARQ feedback delay may render the score $\mathcal S_t$ temporarily inaccurate and trigger a sequence of poor probing decisions and a transient BLER increase. 
To mitigate this behavior, we recommend probing high MCS values only with some probability $p^{\mathrm{probe}}$ if \eqref{eq:probe_condition} holds.\\

 

\subsection{Long-term BLER target enforcement} \label{sec:bler_feedbackloop}

An overly aggressive MCS probing strategy as defined in \eqref{eq:probe_condition} can lead to a transient increase in BLER. 
Moreover, according to \eqref{eq:salad_sinr_update}, the inferred SINR is only mildly decreased following a NACK, which may promote further probing in future transmissions.
To address this, we introduce a feedback mechanism that adapts the instantaneous BLER target $\tau_t$ based on the achieved BLER, inspired by the Proportional-Integral-Derivative (PID) control approach in \cite{nicolini2023link}.

Specifically, we define the integral error
\begin{equation}\label{eq:integral_error}
	E_t := \sum_{i< t-\deltaua_t} \tau - \nack_i
\end{equation}
and adapt the instantaneous BLER target $\tau_t$ as follows:
\begin{equation}
	\tau_t \leftarrow \mathrm{clip} \left( \tau_t + k_E E_t, 0 , 1 \right), \quad k_E>0.
\end{equation}
Observe that if the portion of NACKs exceeds the BLER target $\tau$, $E_t$ becomes negative, resulting in a lower BLER target and a more conservative MCS selection. Conversely, a positive $E_t$ leads to a more aggressive MCS selection. 
This mechanism is instrumental for maintaining the BLER, computed over the whole connection duration, close to the desired target $\tau$.

The full \salad's pseudo-code is outlined in Algorithm \ref{alg:salad}.

\begin{algorithm}
	\caption{\salad: Self-Adaptive Link Adaptation} \label{alg:salad}
	\begin{algorithmic}[1]
		\State \textbf{Parameters} $k_E>0$, $\rho > 0$, $p^{\mathrm{probe}}\in(0,1)$, $\tau^{\mathrm{probe}}>\tau$.
		\State \textbf{Initialize} $E_I=0$, $\sinrest$ arbitrarily.
		\For{slot $t=0,1,\dots$}
		\State Receive set $\mathcal N$ of ACK/NACK. Let $\mathcal U,\mathcal B$ be the corresponding MCS and TBS, respectively.
		\For{$\nack \in \mathcal N$}
		\State Update integral error: $E \leftarrow E + \tau - \nack $
		\EndFor
		
		\Statex \hspace{1em} \texttt{\textcolor{green!60!black}{\# SINR inference update}}
		\For{$(\nack,\mcs, \res) \in (\mathcal N,\mathcal U, \mathcal B)$}
		$$
			\sinrest \leftarrow \sinrest + \varepsilon s^{-1} \left( \blertable(u,\sinrest,b) - \nack \right)
		$$
		\EndFor
	\State Set the latest SINR estimate to $\sinrest_{t-\deltaua_t} := \sinrest$ [\SI{}{\dB}]
		\If{user is scheduled}
     		\Statex \hspace{2.5em} \texttt{\textcolor{green!60!black}{\# SINR underestimation detection}} 
			\State Compute bias score $\mathcal{S}_t$ and its variance $\mathrm{Var}[\mathcal{S}_t | \mathcal H_{0,t} ]$ as in \eqref{eq:score}, \eqref{eq:var_St}, respectively.
			\If{$\tfrac{\mathcal S_t}{\sqrt{\mathrm{Var}[\mathcal{S}_t | \mathcal H_0]}} > \rho \land \mathrm{rand}(0,1)<p^{\mathrm{probe}}$ \label{line:mcs_selection_init}}
			\Statex \hspace{2.5em} \texttt{\textcolor{green!60!black}{\# MCS probing}} 
			\State Set $\tau_t := \tau^{\mathrm{probe}}$
			
			\Else
			
			\State Set $\tau_t := \tau$
			\EndIf
			
			\Statex \hspace{2.5em} \texttt{\textcolor{green!60!black}{\# BLER target $\tau$ enforcement}}
			\State Adjust
			$\tau_t \leftarrow \mathrm{clip} \left( \tau_t \!+ \!k_E E, 0, 1 \right)$
			\Statex \hspace{2.5em} \texttt{\textcolor{green!60!black}{\# MCS selection}} 
			\State Assign MCS $\mcs_t=\mcsilla(\sinrest_{t-\deltaua_t}, \tau_t, \res_t)$ \label{line:mcs_selection_end}

		\EndIf 
		\EndFor
		\State \textbf{Return} $\mcs_t$
	\end{algorithmic}
\end{algorithm}

\subsection{A practical consideration}

For large CBS values, the BLER exhibits a steep transition between its extreme values, 1 and 0, near the Shannon SINR limit. This is illustrated in Table \ref{tab:sigmoid_params}, showing smaller scale parameter $s$ for larger CBS, indicating a steeper BLER cutoff. 

As a result, the term $s_{t-1}^{-1} \left( \blertable_{t-1} - \nack_{t-1} \right)$, which appears in \salad's SINR update \eqref{eq:salad_sinr_update}, can take on extreme values when the TBS is large. Specifically, upon probing, a received ACK leads to a substantial SINR increase $\sinrest_{t}\gg \sinrest_{t-1}$. 
Conversely, receiving a NACK results in only a negligible update $\sinrest_{t+1}\approx \sinrest_t$.
In the absence of probing, the situation is reversed: an ACK leads to a negligible SINR update, while a NACK causes a significant SINR decrease.

A simple and effective remedy is to clip the values of $\blertable_t$ and $s_t$ to appropriate intervals, e.g., $(0.01, 0.99)$ and $(0.5, 10)$, respectively.

\subsection{The impact of delayed HARQ feedback} \label{sec:delayedHARQ}

We conclude the presentation of \salad{} by analyzing the impact of delayed HARQ on SINR inference and MCS selection, while emphasizing its connection to OLLA.\\

\noindent \textbf{Absence of HARQ feedback delay ($\deltaua=0$).} In this case, \salad's SINR inference update can be interpreted as a variant of OLLA with adaptive instantaneous BLER target and offset parameter $\deltaminus$. 
In fact, since the MCS is selected via the up-to-date SINR estimation, i.e., $\mcs_{t-1}=\mcsilla_{t-1}(\sinrest_{t-1}, \tau_t)$, we can approximate the target $\tau_{t-1}$ as the \emph{effective} BLER:
\begin{equation} \label{eq:approx_deltaua0}
	\tau_{t-1} \approx \blertable_{t-1}:= \blertable(u_{t-1},\sinrest_{t-1}), \ \text{if } \delta=0.
\end{equation}
Then, \salad's SINR update \eqref{eq:salad_sinr_update} can be rewritten as
\begin{equation} \label{eq:salad_sinr_update_simple}
	\sinrest_{t} \approx \sinrest_{t-1} + \varepsilon s_{t-1}^{-1} \left( \tau_{t-1} - \nack_{t-1} \right) \quad \forall \, t > 0
\end{equation}
which corresponds to OLLA's update---cfr. with \eqref{eq:olla_update_sa}:
\begin{equation} \label{eq:olla_update_sa_timevar}
	\sinrestolla_t = \sinrestolla_{t-1} + \frac{\deltaminus_{t-1}}{1-\tau_{t-1}} \left( \tau_{t-1} - \nack_{t-1} \right)
\end{equation}
with a time-adaptive offset parameter $\deltaminus_{t-1} := \varepsilon s^{-1}_{t-1} (1 - \tau_{t-1})$ and a dynamic target $\tau_{t-1}$. \\

\noindent \textbf{Presence of HARQ feedback delay ($\deltaua>0$).} In this scenario, \salad's SINR update differs significantly from that of time-adaptive OLLA. In both algorithms, the MCS $\mcs_{t-1}$ is selected based on an \emph{outdated} SINR estimation, specifically, $\mcs_{t-1}=\mcsilla_{t-1}(\sinrest_{t-1-\deltaua_{t-1}}, \tau_{t-1})$. 
Yet, the SINR staleness renders \eqref{eq:approx_deltaua0} invalid: the target $\tau_{t-1}$, used by OLLA in its SINR update, is different from the \emph{effective} BLER, $ \blertable_{t-1}$, used by \salad{}. In other words, \salad{} updates the SINR using fresher information than OLLA.
As a result, \salad's SINR estimation is more robust to feedback delay, as confirmed by the experiments in Section~\ref{sec:experiments_controlled_gain}.


\section{Experimental Results} \label{sec:results}

To evaluate \salad's performance, we conduct full-stack system simulations and real-world over-the-air (OTA) measurements using a real-time 5G testbed. 
Specifically, we implement \salad{}\footnote{A slight variation of \salad{} is tested, where the instantaneous BLER target is adjusted via integral error (line 21) only if probing is not performed.} and OLLA in the 5G downlink (PDSCH) scheduler within the gNB multiple access (MAC) layer of the OpenAirInterface (OAI) \cite{kaltenberger2025driving} using the Sionna Research Kit (SRK) \cite{cammerer2025sionna}. 
Using OAI's radio frequency simulator (RFSim), we perform full-stack simulations aligned with the 5G NR standard.
Furthermore, leveraging the NVIDIA Aerial CUDA-accelerated RAN platform, we carry out real-world OTA measurements with a commercial off-the-shelf (COTS) Open RAN radio unit (O-RU) and a COTS user equipment (UE) deployed inside an anechoic chamber. All experiments in this section adopt a long-term BLER target value of $\tau=0.1$, a widely used setting in the literature \cite{wu2011coding,park2015optimizing}. \\

\begin{table}
	\caption{\salad{}'s configuration parameters. Default parameters are designed manually. The other two configurations are optimized for a specific low/high throughput regime by Nelder-Mead method.}
	\label{tab:salad_config}
	\centering
	\begin{tabular}{@{}l l l l@{}}
		\toprule[1pt]
		\textbf{Parameter} & \shortstack[l]{\textbf{Default} \\ \textbf{(manual design)}} & \shortstack[l]{\textbf{Optimized for} \\ \textbf{low throughput}} & \shortstack[l]{\textbf{Optimized for} \\ \textbf{high throughput}} \\
		\midrule[0.5pt]
		$\varepsilon$ & 1 & 1.09 & 0.85 \\
		$\rho$ & 0.25 & 0.335 & 0.346 \\
		$T$ & 15 & 13 & 13 \\
		$p^{\mathrm{probe}}$ & 0.15 & 0.134 & 0.139 \\
		$\tau^{\mathrm{probe}}$ & 0.999 & 0.941 & 0.957 \\
		$k_E$ & 0.01 & 0.032 & 0.0017 \\
		\bottomrule[1pt]
	\end{tabular}
\end{table}

\begin{figure*}[t]
	\centering
	\begin{subfigure}[t]{0.48\textwidth}
		\centering
		\includegraphics[width=\textwidth]{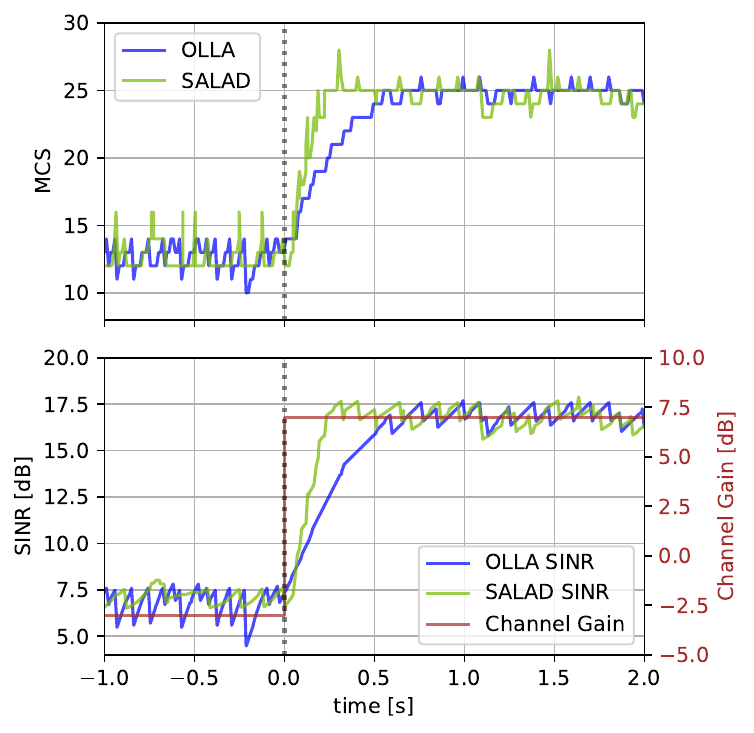} 
		\caption{\textbf{Channel gain surge}: No CQI feedback, SISO with 1 O-RU antenna, 24 PRBs.}
		\label{fig:rfsim_siso_channel_surge}
	\end{subfigure}
	\hfill
	\begin{subfigure}[t]{0.48\textwidth}
		\centering
		\includegraphics[width=\textwidth]{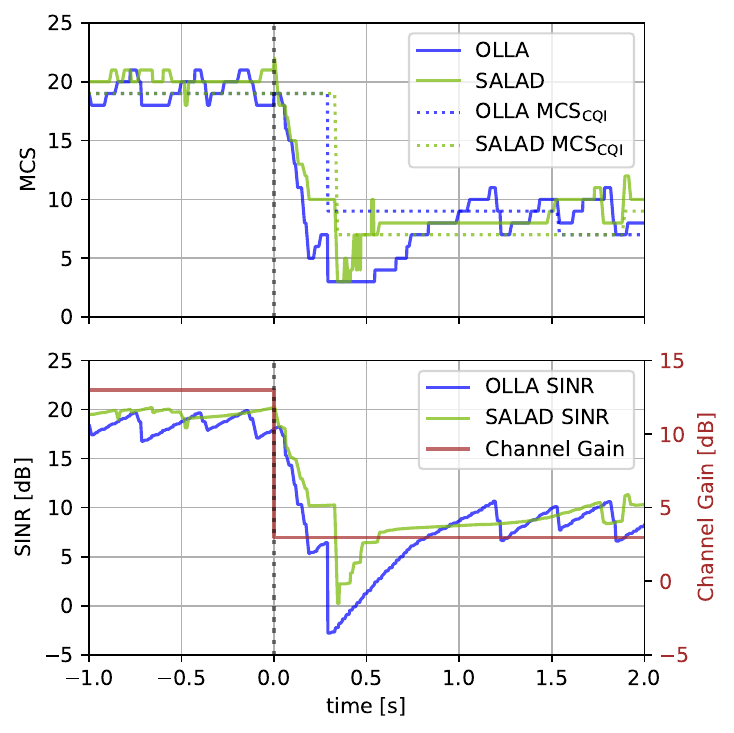}
		\caption{\textbf{Channel gain drop}: With CQI feedback, single-layer MISO with 2 O-RU antennas, 51 PRBs.}
		\label{fig:rfsim_csi-rs_channel_drop}
	\end{subfigure}
	\caption{\textbf{Controlled SINR.} SINR estimation and MCS index for \salad{} and OLLA from OAI RFSim with 7DS2U TDD slot pattern.}
\end{figure*}

\subsection{5G testbed with controlled SINR} \label{sec:experiments_controlled_gain}

We first evaluate \salad's performance via full-stack, system-level simulations using OAI's RF simulator (RFSim). 
The pathloss in RFSim is controlled via a testbench script that alters the channel gain at predefined timestamps.
The script also loads application-layer traffic from an external traffic generator server (\texttt{iperf3 -s}) to a traffic generator app (\texttt{iperf3}) running on OAI's software-defined UE, which is connected to the software-defined gNB via RFSim.
We simulate two RF channel configurations: (i) a single-input single-output (SISO) AWGN channel without CQI feedback, i.e., CSI-RS was disabled, and (ii) a multiple-input single-output (MISO) AWGN channel with CQI feedback.
For the frame structure, we use the 5G NR N78 band with a 7DS2U (7 DL slots, 1 special slot, 2 UL slots) TDD slot pattern. 
RFSim does not run in real-time and advances according to simulation speed. 
Yet, higher protocol layers, e.g., MAC and above, run in real system time and are 5G-NR-compliant.
SALAD's default parameters (see Table \ref{tab:salad_config}) are used. \\

\noindent \textbf{Channel surge}. Figure~\ref{fig:rfsim_siso_channel_surge} shows the time evolution of the scheduled MCS indices for \salad{} and OLLA (top), along with their corresponding SINR estimates and the channel gain (bottom). The channel gain surges from \SI{-3}{\dB} to \SI{7}{\dB} at timestamp $t=0$.  The UDP downlink throughput is \SI{1}{\Mbps}. 
Figure~\ref{fig:rfsim_siso_channel_surge} shows that SALAD adapts to the improved channel conditions approximately twice as fast as OLLA (with $\Delta^{\text{NACK}}=1$), thanks to better SINR inference enabled by MCS probing. 
In fact, upon the channel gain surge, \salad{} probes high MCS indices (top) which translate to better SINR inference (bottom). 
However, \salad{} also probes high MCS values when unnecessary (top), particularly before the surge. This can be mitigated by decreasing the probing probability $p^{\mathrm{probe}}$ and/or increasing the calibration ratio $\rho$. \\




\noindent \textbf{Channel drop, delayed feedback}. We enable CSI-RS and use a variant of \salad{} that incorporates CQI reports $\sinrrep$ by simply considering $\sinrrep+\sinrest$ as the current SINR estimate, similarly to OLLA. 
Figure~\ref{fig:rfsim_csi-rs_channel_drop} shows the time evolution of the scheduled MCS indices of \salad{} and OLLA (solid lines, top) along with the corresponding MCS derived from the reported CQI ($\mathrm{MCS}_{\mathrm{CQI}}$) \cite{3gpp.38.214}, as well as their SINR estimates and the channel gain (bottom). The channel gain drops from \SI{13}{\dB} to \SI{3}{\dB} at time $t=0$. As before, the DL UDP throughput is \SI{1}{\Mbps}. 
We observe from Figure~\ref{fig:rfsim_csi-rs_channel_drop} that, immediately after the channel degradation, \salad{} and OLLA (with $\Delta^{\text{NACK}}=1$) decrease the MCS at a similar rate. 
However, the delayed CQI report, received at time $t\approx\,$\SI{0.3}{\sec}, causes the estimated SINR to drop further, resulting in an undershoot in both SINR and MCS for both algorithms. 
\salad{} promptly compensates for this undershoot by resuming the probing of higher MCS values, enabling a faster recovery compared to OLLA. 

This behavior stems from Section~\ref{sec:delayedHARQ}: \salad{} is robust to feedback inaccuracies since it updates its SINR according to the \emph{effective} BLER which, under delayed feedback, may considerably differ from the BLER target considered by OLLA.



\subsection{Over-the-air experiments}

Next, we benchmark the performance of SALAD against that of OLLA in a real-time over-the-air (OTA) 5G testbed. 
Our implementation leverages the Aerial RAN CoLab Over-the-Air (ARC-OTA) platform \cite{nvidiaAerial}, which integrates NVIDIA’s Aerial CUDA-Accelerated RAN (using NVIDIA cuPHY Layer 1 \cite{cuphy}) with OAI’s higher protocol stacks \cite{kaltenberger2025driving}.

\begin{figure*}[t]
	\centering
	\includegraphics[height=4.85cm]{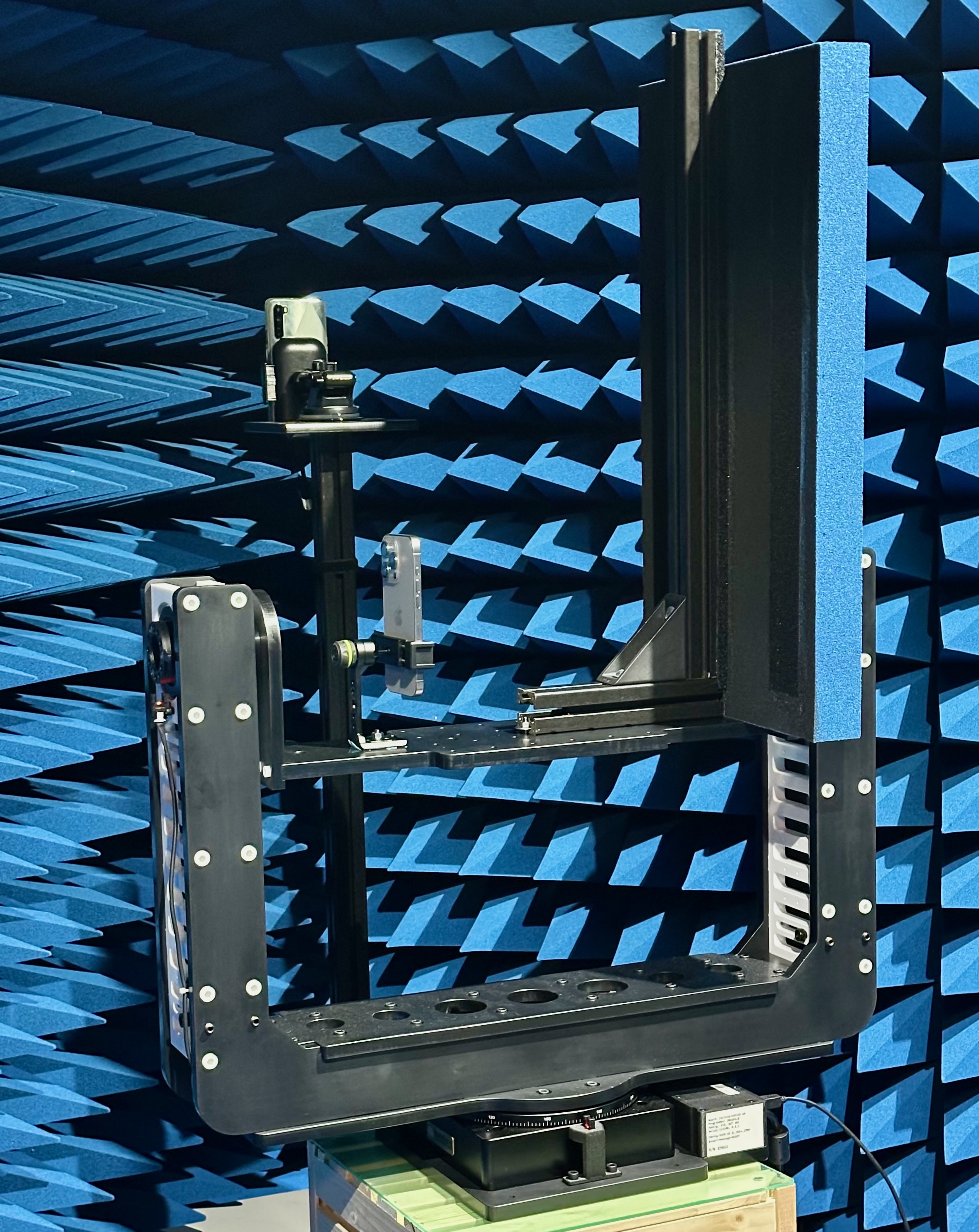}
	\hfill
	\includegraphics[height=4.85cm]{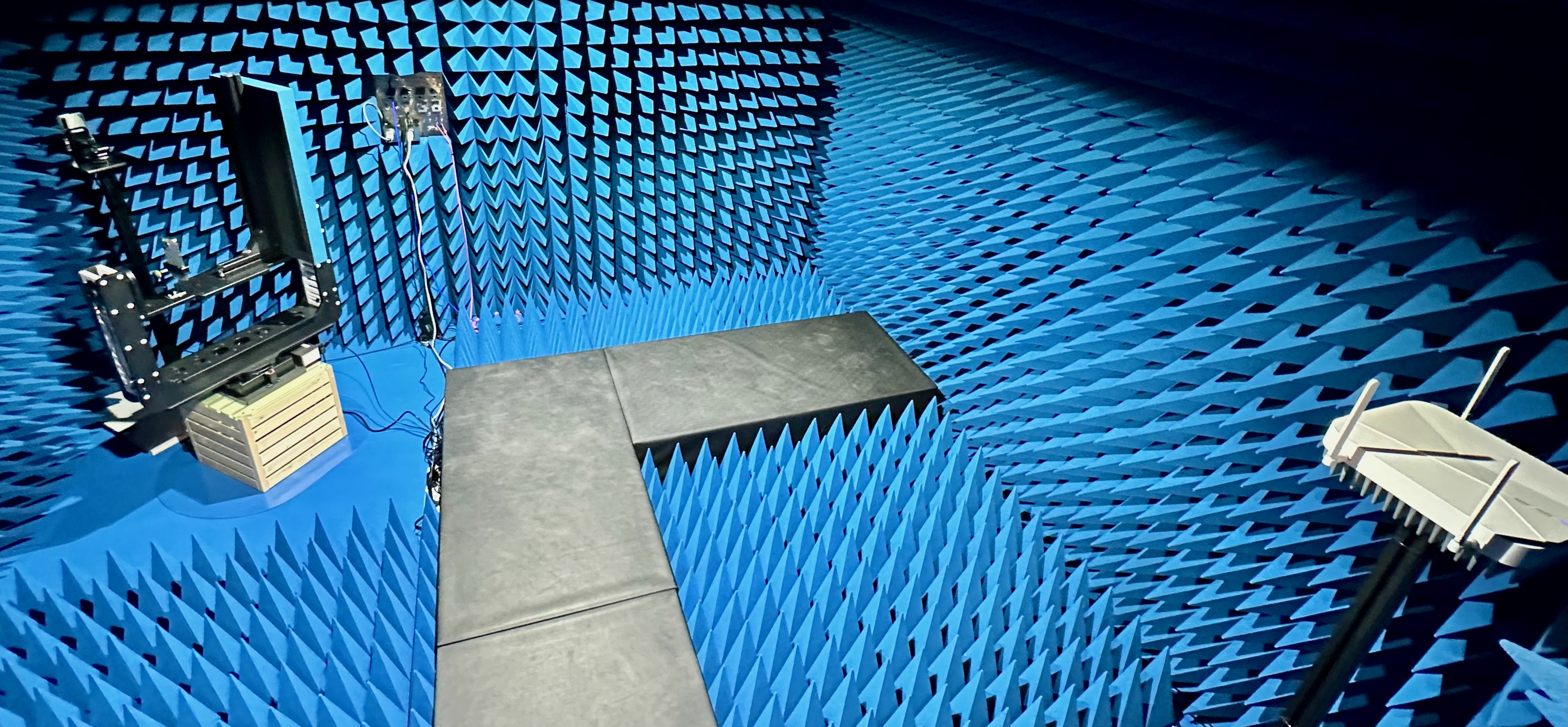}
	\hfill
	\includegraphics[height=4.85cm]{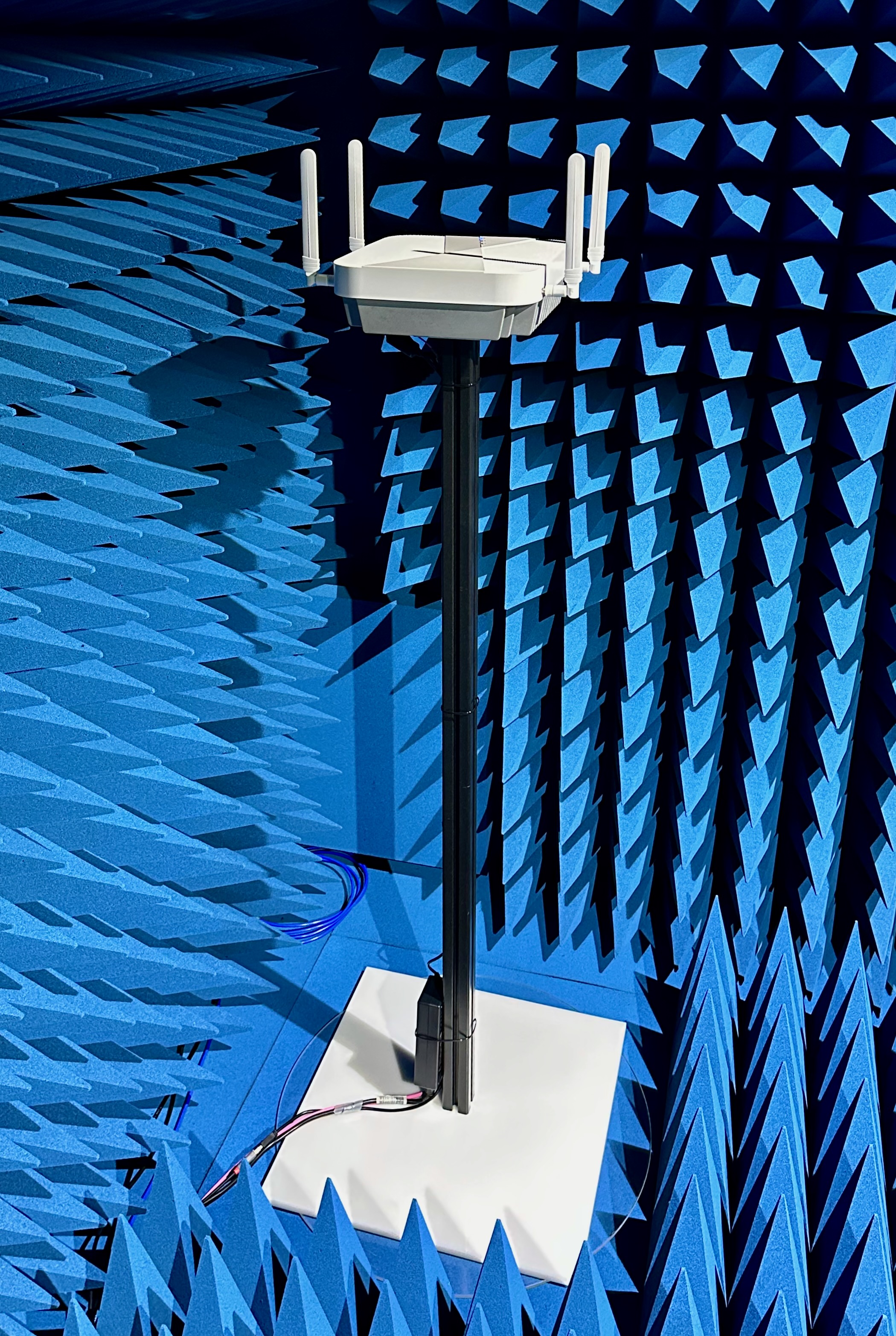}
	\caption{Measurement setup in anechoic chamber at ETH Z\"urich with COTS UE and rotational absorber table (left) and COTS O-RU (right).}
	\label{fig:anechoic_chamber}
\end{figure*}

Figure~\ref{fig:anechoic_chamber} shows our controllable radio environment inside an anechoic chamber. Two UEs are placed in the left corner, one behind and one above a rotational table. A flat RF absorber panel is mounted on the opposite side of the rotational table, facing the UEs.
The table is shown in Figure~\ref{fig:anechoic_chamber} at a reference rotation angle of $\phi=\SI{280}{\degree}$.  
Depending on the table's rotation angle, the UEs experience either line-of-sight (LoS) or non-line-of-sight (NLoS) conditions with respect to the O-RU. 
In the right corner, we place a COTS O-RU with 4 antennas and operating with a \SI{100}{\MHz} channel bandwidth (273 PRBs) in the 5G N78 band. 
The O-RU's transmit power is set to \SI{9}{\dBm} per antenna to prevent MCS saturation in LoS while maintaining connectivity in NLoS. \\

\noindent \textbf{Experimental setup}. A script controls the table rotation while recording the rotation angles and the corresponding timestamps. 
The table is equipped with a high-precision rotation axis motor which is re-calibrated after each power cycle.
This setup ensures that the RF channel characteristics remain consistent throughout our measurement campaign.
We use a COTS UE (OnePlus Nord) and rotate the rotational table back and forth. The rotation begins in a LoS position ($\phi=\SI{100}{\degree}$, i.e., the absorber's location is opposite to its placement in Figure~\ref{fig:anechoic_chamber}) and proceeds at constant angular speed until the UE is in NLoS position ($\phi=\SI{190}{\degree}$, with the absorber closest to the UE). 
Then, the table rotates back to the starting position. The rotation speed is set to \SI{1300}{\degree\per\minute}, resulting in a full back-and-forth cycle lasting \SI{8}{\sec}. 
CQI reports are observed to be inaccurate and unreliable. For this reason, CQI reports are disabled for our OTA experiments.\\

\begin{figure}
	\centering
	\includegraphics[width=\columnwidth]{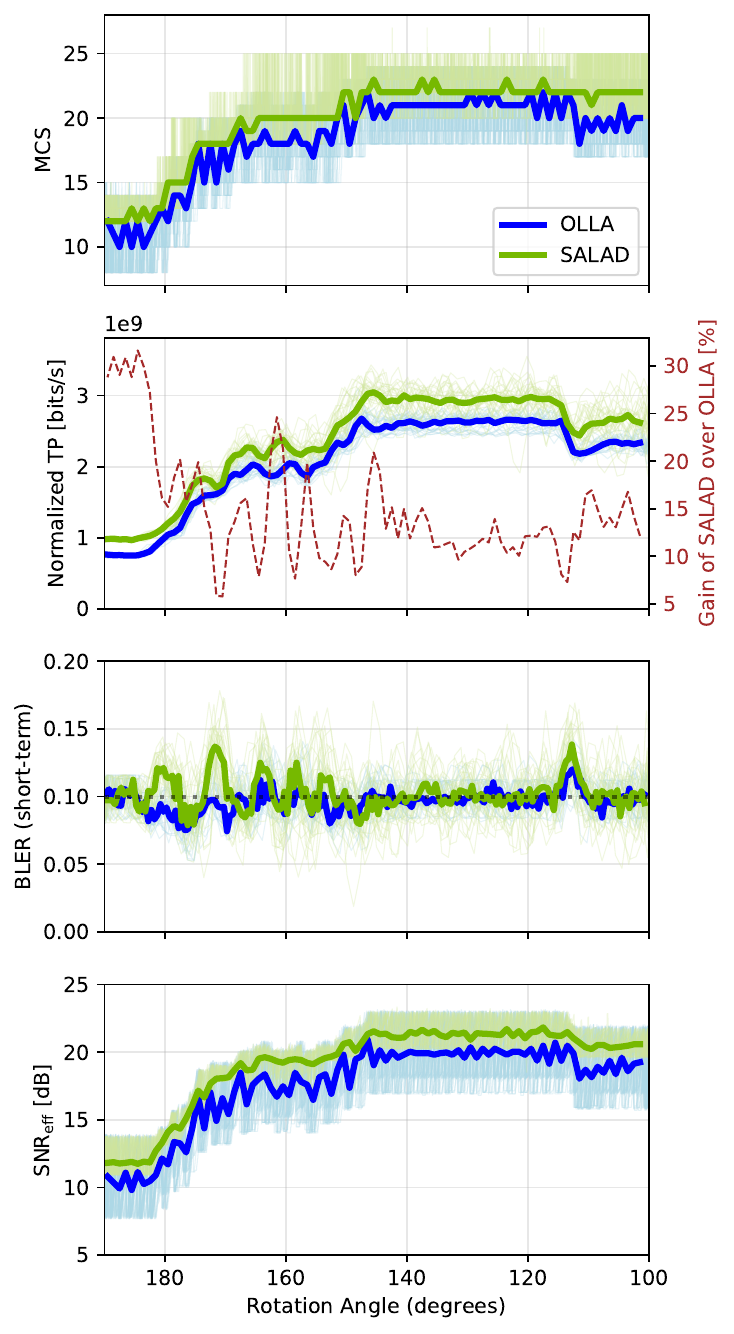}
	\caption{\textbf{OTA---Default SALAD and OLLA}: Default \salad{} (see Table \ref{tab:salad_config}) vs. default OLLA ($\deltaminus=1$) in improving channel conditions (NLoS to LoS) with maximum TCP throughput, single-layer MIMO, 4 O-RU antennas, 273 PRBs, 3DSU TDD slot pattern.}
	\label{fig:over_the_air_results}
\end{figure}

\noindent \textbf{Default SALAD vs. default OLLA}. Figure~\ref{fig:over_the_air_results} provides a first visual comparison of the performance of \salad{} and OLLA in OTA experiments, shown as a function of rotation angle. \salad{} uses its manually designed ``default'' configuration (see Table \ref{tab:salad_config}), while OLLA employs an offset update $\Delta^{\text{NACK}}=1$, which is regarded as the default value in the literature, see \cite{peralta2022outer,saxena2020bayesian}. 
All experiments are obtained under identical controlled RF conditions with a TCP throughput reaching \SI{180}{\Mbps} in LoS and \SI{70}{\Mbps} in NLoS. 

Figure~\ref{fig:over_the_air_results} superposes the results from 30 rotation trajectories. 
The individual trajectories are then overlayed and shown in light shading. The median of all trajectory data falling within the same angle bin is plotted in dark thick lines.
For clarity of visualization, Figure~\ref{fig:over_the_air_results} focuses exclusively on the backward trajectory, i.e., from NLoS to LoS.

The upper subplot in Figure~\ref{fig:over_the_air_results} shows the MCS selected by \salad{} and OLLA. \salad{} consistently selects higher MCS values and better tracks the channel gain surge, whereas OLLA tends to oscillate more under stable channel conditions, resulting in efficiency loss with respect to \salad{}.
The second subplot in Figure~\ref{fig:over_the_air_results} shows the ``normalized'' throughput (TP)
\begin{equation}
	\text{Normalized TP} = \frac{\text{TP}_{\text{1st-round}}\times \text{SE}_{\text{sched}}}{\text{SE}_{\text{min}}},
\end{equation}
defined as the throughput achieved during the first HARQ round ($\text{TP}_{\text{1st-round}}$), i.e., computed only for packets which do not require retransmission, multiplied by the SE of the scheduled MCS ($\text{SE}_{\text{sched}}$) and scaled by the minimum available SE ($\text{SE}_{\text{min}}$).
This performance metric jointly captures SE, raw connection speed, and reliability. It is more practically relevant than the classic expected SE formulation in \eqref{eq:pb}.
\salad{} consistently achieves higher normalized throughput than OLLA with gains (in orange) reaching up to 30\%.
Indeed, selecting a higher MCS and achieving a higher instantaneous rate can enable the TCP congestion control mechanism to further increase its future transmission rate, thereby creating a virtuous circle.

Although not visualized here, both \salad{} and OLLA can maintain the long-term BLER, computed over the entire connection duration, at the target value $\tau=0.1$. 
The third subplot in Figure~\ref{fig:over_the_air_results} presents the short-term BLER, computed over a sliding window of approximately \SI{250}{\milli\s}. For both algorithms, the short-term BLER fluctuates around the target $\tau=0.1$. SALAD exhibits a slight BLER over-shoot (approximately $0.135$ versus $\tau=0.1$) during the channel gain surge, at an angle range of \SI{170}{\degree}-\SI{175}{\degree}, due to aggressive MCS probing. In contrast, OLLA's BLER undershoots during the same interval, indicating slower adaptation to improving channel conditions. 
The second subplot shows that \salad's BLER overshoot is compensated by improved overall performance. 

The lower subplot in Figure~\ref{fig:over_the_air_results} shows the SINR $\sinrest$ estimated by \salad{} and OLLA. \salad's SINR estimate is consistently higher and more stable than OLLA's. 
Interestingly, even when the short-term BLER overshoots, the SINR estimation remains unaffected by the received NACKs. 
This stems directly from expression \eqref{eq:salad_sinr_update}: when the predicted BLER is high and a NACK is received, the estimated SINR is only slightly decreased. In contrast, OLLA always penalizes NACKs with a fixed SINR decrement of $\deltaminus$. \\

\begin{figure}[htbp]
	\centering
	
	\begin{subfigure}[b]{.5\textwidth}
		\centering
		\begin{subfigure}[b]{0.38\textwidth}
			\includegraphics[width=\textwidth]{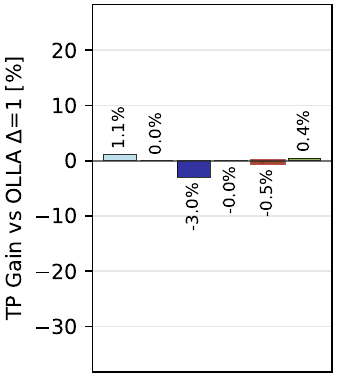}
			\caption*{(a.1) \SI{1}{\Mbps}}
			\phantomsubcaption\label{fig:customA}
		\end{subfigure}
		\hfill
		\begin{subfigure}[b]{0.28\textwidth}
			\includegraphics[width=\textwidth]{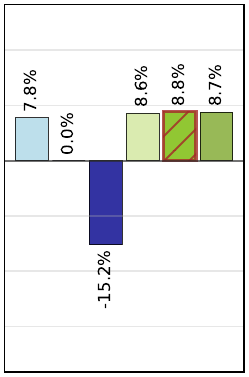}
			\caption*{(a.2) \SI{10}{\Mbps}}
		\end{subfigure}
		\hfill
		\begin{subfigure}[b]{0.28\textwidth}
			\includegraphics[width=\textwidth]{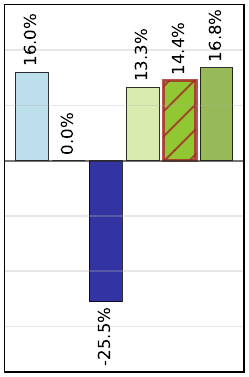}
			\caption{(a.3) \SI{100}{\Mbps}}
		\end{subfigure}
		\caption*{(a) \textbf{Normalized throughput gain vs. OLLA ($\deltaminus=1$)}.}
		\label{fig:top_group}
	\end{subfigure}
	
	\vspace{0.8cm} 
	
	\begin{subfigure}[b]{.5\textwidth}
		\centering
		\begin{subfigure}[b]{0.375\textwidth}
			\includegraphics[width=\textwidth]{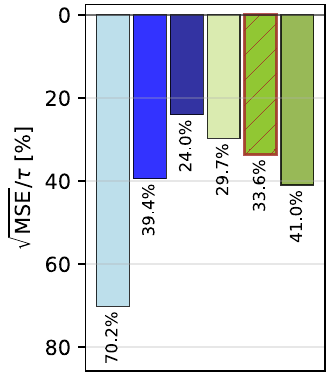}
			\caption*{(b.1) \SI{1}{\Mbps}}
		\end{subfigure}
		\hfill
		\begin{subfigure}[b]{0.28\textwidth}
			\includegraphics[width=\textwidth]{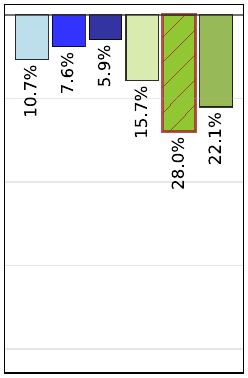}
			\caption*{(b.2) \SI{10}{\Mbps}}
		\end{subfigure}
		\hfill
		\begin{subfigure}[b]{0.28\textwidth}
			\includegraphics[width=\textwidth]{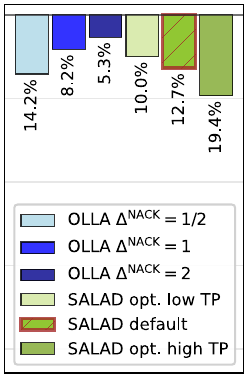}
			\caption*{(b.3) \SI{100}{\Mbps}}
		\end{subfigure}
		\caption*{(b) \textbf{Short-term BLER deviation from long-term target $\tau$}.}
		\label{fig:bottom_group}
	\end{subfigure}
	
	\caption{\textbf{OTA---Variants of \salad{} and OLLA}: Performance of various OLLA variants and SALAD with default and optimized parameters, comparing normalized TP (a) and standard error to the BLER target (b) for three different application-layer TP scenarios: \SI{1}{\Mbps} UDP, \SI{10}{\Mbps} UDP, and \SI{100}{\Mbps} TCP.}
	\label{fig:pareto_frontier}
\end{figure}

\noindent \textbf{One SALAD vs. many OLLAs}. One may wonder if OLLA can match \salad's performance with proper tuning. 
Figure~\ref{fig:pareto_frontier} shows that a \emph{single} \salad{} configuration can outperform or match \emph{each} OLLA instance across a range of scenarios. 
In other words, OLLA would require proper dynamic parameter tuning to approach \salad's performance, while \salad{} can perform well out-of-the-box in different scenarios.
Specifically, we consider three traffic regimes: (i) constant \SI{1}{\Mbps} UDP traffic, (ii) constant \SI{10}{\Mbps} UDP traffic, and (iii) highest possible TCP throughput, averaging \SI{100}{\Mbps}. 
For each traffic regime, we report the achieved normalized throughput (top subplot) relative to default OLLA ($\deltaminus=1$) and the root mean square deviation of the instantaneous BLER from the target $\tau=0.1$ (bottom), expressed as a percentage of the target $\tau$. A lower value of the metric indicates that the BLER stays closer to the target at all times, i.e., the algorithm can promptly adapt to varying channel conditions.
Both metrics are computed as trajectory mean over the full rotation cycle. 

To interpret Figure~\ref{fig:pareto_frontier}, we first examine the manually designed ``default'' \salad{} configuration (see Table~\ref{tab:salad_config}).
Default \salad{} is compared with three OLLA variants: ``slow'' OLLA with NACK offset $\deltaminus=0.5$, default OLLA ($\deltaminus=1$), and ``fast'' OLLA ($\deltaminus=2$).
In the low-throughput regime (\SI{1}{\Mbps}), all algorithms achieve similar normalized throughput, but ``slow'' OLLA fails to meet the BLER target. In fact, when the throughput is low, the frequency of HARQ reports is also low, making channel variations appear rapid from the perspective of a link adaptation algorithm that updates at every HARQ feedback. This renders ``slow'' OLLA unable to keep up with channel variations. 
In the mid-throughput regime (\SI{10}{\Mbps}), all algorithms can closely meet the BLER target. Yet, in terms of throughput, \salad{} performs on par with ``slow'' OLLA and outperforms both default and ``fast'' OLLA by 8.8\% and an additional 15.2\%, respectively. 
In the high-throughput regime (\SI{100}{\Mbps}), default SALAD again matches ``slow'' OLLA and outperforms the other OLLA variants, achieving throughput gains of 14.4\% over default OLLA and an additional 25.5\% over ``fast'' OLLA. 
In fact, the high frequency of HARQ feedback makes channel variations appear slow, favoring more conservative adaptation.

As a result, \salad{} rightly earns the ``self-adaptation'' label, since it exhibits customized behavior tailored to each situation, without any dynamic parameter tuning. \\


\noindent \textbf{\salad's parameter tuning}.
Finally, we investigate the extent to which \salad's parameters can be customized in a specific scenario, and how this may negatively impact \salad's performance in different traffic regimes.
To this end, \salad's parameters are optimized in the low throughput scenario (denoted as ``opt. low TP'' in Figure~\ref{fig:pareto_frontier}) and the high throughput scenario (``opt. high TP'') separately. 

For parameter optimization we use the Nelder-Mead algorithm \cite{nelder1965simplex}, which (i) does not require any gradient information and (ii) is well-suited for objective functions that are expensive to evaluate \cite{fletcher2000practical}. 
Such features suit our needs: (i) the objective, defined as a linear combination of normalized throughput and the negative mean squared deviation of the BLER from the target $\tau$, is not available in closed-form. 
Furthermore, (ii) evaluating the objective for one parameter configuration requires multiple full table rotations, lasting several tens of seconds overall, with an additional settling time of a few seconds between consecutive trials to allow transient effects to vanish. 
In each traffic regime, the initial configuration is set to the ``default'' one, and Nelder-Mead iterates over 25 different steps. The same process is repeated three times in the low throughput regime and five times in high-throughput.
The resulting optimal parameters are reported in Table \ref{tab:salad_config}.

Finally, we evaluate the performance of both fine-tuned configuration across all traffic regimes (low, mid and high throughput), as shown in Figure~\ref{fig:pareto_frontier}.
As expected, the fine-tuned \salad{} configuration outperforms the other \salad's variants in its preferred regime, although the gains remains limited. Moreover, the default parameters, manually selected prior to the optimization campaign, turns out to be a reasonable choice.
This demonstrates that \salad's performance is quite stable across a range of parameter settings.

\section{Conclusions}

In this paper, we have introduced \salad{}, a novel algorithm for adaptive MCS selection for wireless communications.
\salad{} infers the user's effective SINR from ACK/NACK feedback alone by minimizing a regularized cross-entropy loss with respect to predicted BLER values using gradient descent.
To accelerate MCS adaptation under improving channel conditions, \salad{} leverages hypothesis testing to detect SINR underestimation and safely probes high MCS values.
This improves both short-term spectral efficiency and long-term SINR inference accuracy.
A feedback control loop ensures that the long-term BLER remains close to the desired target.

We have evaluated \salad{} in OTA experiments and benchmarked it against the industry-standard OLLA. 
Results demonstrated that \salad{} delivers robust and consistent performance across a wide range of traffic regimes, achieving up to 15\% gains in throughput and spectral efficiency over default OLLA configurations while reliably meeting the BLER target.
Additionally, \salad{} can be fine-tuned to specific traffic regimes using derivative-free optimization techniques.

A natural extension of this work involves learning an optimal MCS selection rule conditioned on the latest SINR estimate, bias score, and integral error, thereby replacing the heuristic in lines \ref{line:mcs_selection_init}–\ref{line:mcs_selection_end} of Algorithm~\ref{alg:salad}. This approach may be extended to handle CQI reports more effectively.

\balance 
\bibliographystyle{IEEEtran}
\bibliography{literature}

@inproceedings{sampath1997setting,
	title={{On setting reverse link target SIR in a CDMA system}},
	author={Sampath, Ashwin and Kumar, P Sarath and Holtzman, Jack M},
	booktitle={1997 IEEE 47th Vehicular Technology Conference. Technology in Motion},
	volume={2},
	pages={929--933},
	year={1997},
	organization={IEEE}
}

@inproceedings{pedersen2007frequency,
	title={{Frequency domain scheduling for OFDMA with limited and noisy channel feedback}},
	author={Pedersen, Klaus I and Monghal, Guillaume and Kovacs, Istvan Z and Kolding, Troels E and Pokhariyal, Akhilesh and Frederiksen, Frank and Mogensen, Preben},
	booktitle={2007 IEEE 66th Vehicular Technology Conference},
	pages={1792--1796},
	year={2007},
	organization={IEEE}
}

@inproceedings{delgado2017fast,
	title={Fast convergence outer loop link adaptation with infrequent updates in steady state},
	author={Delgado, Ramon A and Lau, Katrina and Middleton, Richard and Karlsson, Robert S and Wigren, Torbjoern and Sun, Ying},
	booktitle={2017 IEEE 86th vehicular technology conference (VTC-Fall)},
	pages={1--5},
	year={2017},
	organization={IEEE}
}

@inproceedings{kela2022reinforcement,
	title={Reinforcement learning for delay sensitive uplink outer-loop link adaptation},
	author={Kela, Petteri and H{\"o}hne, Thomas and Veijalainen, Teemu and Abdulrahman, Hussein},
	booktitle={2022 Joint European Conference on Networks and Communications \& 6G Summit (EuCNC/6G Summit)},
	pages={59--64},
	year={2022},
	organization={IEEE}
}

@inproceedings{mandelli2021troll,
	title={{TROLL: Training of outer loop link adaptation in wireless networks via back-propagation}},
	author={Mandelli, Silvio and Weber, Andreas and Baracca, Paolo and Mohammadi, Jafar},
	booktitle={WSA 2021; 25th International ITG Workshop on Smart Antennas},
	pages={1--6},
	year={2021},
	organization={VDE}
}

@inproceedings{saxena2020bayesian,
	title={{Bayesian link adaptation under a BLER target}},
	author={Saxena, Vidit and Jald{\'e}n, Joakim},
	booktitle={2020 IEEE 21st International Workshop on Signal Processing Advances in Wireless Communications (SPAWC)},
	pages={1--5},
	year={2020},
	organization={IEEE}
}

@inproceedings{park2015optimizing,
	title={Optimizing the target error rate for link adaptation},
	author={Park, Sungwoo and Daniels, Robert C and Heath, Robert W},
	booktitle={2015 IEEE Global Communications Conference (GLOBECOM)},
	pages={1--6},
	year={2015},
	organization={IEEE}
}

@article{saxena2021reinforcement,
	title={Reinforcement learning for efficient and tuning-free link adaptation},
	author={Saxena, Vidit and Tullberg, Hugo and Jald{\'e}n, Joakim},
	journal={IEEE Transactions on Wireless Communications},
	volume={21},
	number={2},
	pages={768--780},
	year={2021},
	publisher={IEEE}
}

@inproceedings{saxena2019contextual,
	title={Contextual multi-armed bandits for link adaptation in cellular networks},
	author={Saxena, Vidit and Jald{\'e}n, Joakim and Gonzalez, Joseph E and Bengtsson, Mats and Tullberg, Hugo and Stoica, Ion},
	booktitle={Proceedings of the 2019 Workshop on Network Meets AI \& ML},
	pages={44--49},
	year={2019}
}

@article{robbins1951stochastic,
	title={A stochastic approximation method},
	author={Robbins, Herbert and Monro, Sutton},
	journal={The annals of mathematical statistics},
	pages={400--407},
	year={1951},
	publisher={JSTOR}
}

@article{blum1954approximation,
	title={Approximation methods which converge with probability one},
	author={Blum, Julius R},
	journal={The Annals of Mathematical Statistics},
	pages={382--386},
	year={1954},
	publisher={JSTOR}
}

@article{dieuleveut2020bridging,
	author    = {Aymeric Dieuleveut and Alain Durmus and Francis Bach},
	title     = {{Bridging the gap between constant step size stochastic gradient descent and Markov chains}},
	journal   = {Annals of Statistics},
	volume    = {48},
	number    = {3},
	pages     = {1348--1382},
	year      = {2020},
	month     = {June},
}

@software{sionna,
	title = {Sionna},
	author = {Hoydis, Jakob and Cammerer, Sebastian and {Ait Aoudia}, Fayçal and
	Nimier-David, Merlin and Maggi, Lorenzo and Marcus, Guillermo and Vem, Avinash and Keller,
	Alexander},
	note = {https://nvlabs.github.io/sionna/},
	year = {2022},
	version = {1.1.0}
}

@techreport{3gpp.38.214,
	title        = {{3rd Generation Partnership Project; Technical Specification Group Radio Access Network; NR; Physical layer procedures for data (3GPP TS 38.214)}},
	institution  = {3GPP},
	number       = {TS 38.214},
	version      = {18.2.0},
	year         = {2024},
	month        = {May},
	note         = {Release 18},
}

@inproceedings{combes2014optimal,
	title={Optimal rate sampling in 802.11 systems},
	author={Combes, Richard and Proutiere, Alexandre and Yun, Donggyu and Ok, Jungseul and Yi, Yung},
	booktitle={IEEE INFOCOM 2014-IEEE Conference on Computer Communications},
	pages={2760--2767},
	year={2014},
	organization={IEEE}
}

@inproceedings{nicolini2023link,
	title={{Link Adaptation Algorithm for Optimal Modulation and Coding Selection in 5G and Beyond Systems}},
	author={Nicolini, Andrea and Icolari, Vincenzo and Dardari, Davide},
	booktitle={ICC 2023-IEEE International Conference on Communications},
	pages={5279--5284},
	year={2023},
	organization={IEEE}
}

@article{peri2024offline,
	title={{Offline Reinforcement Learning and Sequence Modeling for Downlink Link Adaptation}},
	author={Peri, Samuele and Russo, Alessio and Fodor, Gabor and Soldati, Pablo},
	journal={arXiv preprint arXiv:2410.23031},
	year={2024}
}

@article{wu2011coding,
	title={{Coding versus ARQ in fading channels: How reliable should the PHY be?}},
	author={Wu, Peng and Jindal, Nihar},
	journal={IEEE Transactions on Communications},
	volume={59},
	number={12},
	pages={3363--3374},
	year={2011},
	publisher={IEEE}
}

@article{martin2021emerging,
	title={{Emerging tools for link adaptation on 5G NR and beyond: Challenges and opportunities}},
	author={Mart{\'\i}n-Vega, Francisco J and Ruiz-Sicilia, Juan Carlos and Aguayo, Mari Carmen and G{\'o}mez, Gerardo},
	journal={IEEE Access},
	volume={9},
	pages={126976--126987},
	year={2021},
	publisher={IEEE}
}

@article{kamerman1997wavelan,
	title={{WaveLAN{\textregistered}-II: a high-performance wireless LAN for the unlicensed band}},
	author={Kamerman, Ad and Monteban, Leo},
	journal={Bell Labs technical journal},
	volume={2},
	number={3},
	pages={118--133},
	year={1997},
	publisher={Wiley Online Library}
}

@phdthesis{bicket2005bit,
	title={Bit-rate selection in wireless networks},
	author={Bicket, John Charles},
	year={2005},
	school={Massachusetts Institute of Technology}
}

@inproceedings{nakamura2002adaptive,
	title={{Adaptive control of link adaptation for high speed downlink packet access (HSDPA) in W-CDMA}},
	author={Nakamura, Michiharu and Awad, Yassin and Vadgama, Sunil},
	booktitle={The 5th International Symposium on Wireless Personal Multimedia Communications},
	volume={2},
	pages={382--386},
	year={2002},
	organization={IEEE}
}

@article{buenestado2014analysis,
	title={{Analysis of throughput performance statistics for benchmarking LTE networks}},
	author={Buenestado, V{\'\i}ctor and Ruiz-Aviles, Jos{\'e} M and Toril, Matias and Luna-Ram{\'\i}rez, Salvador and Mendo, Adriano},
	journal={IEEE Communications letters},
	volume={18},
	number={9},
	pages={1607--1610},
	year={2014},
	publisher={IEEE}
}

@inproceedings{zhu2023nolla,
	title={{NOLLA: Non-Linear Outer Loop Link Adaptation for Enhancing Wireless Link Transmission}},
	author={Zhu, Lingrui and Bockelmann, Carsten and Schier, Thorsten and Hajri, Salah Eddine and Dekorsy, Armin},
	booktitle={2023 IEEE 34th Annual International Symposium on Personal, Indoor and Mobile Radio Communications (PIMRC)},
	pages={1--6},
	year={2023},
	organization={IEEE}
}

@article{duran2015self,
	title={{Self-optimization algorithm for outer loop link adaptation in LTE}},
	author={Duran, A and Toril, Mat{\'\i}as and Ruiz, Fernando and Mendo, Adriano},
	journal={IEEE Communications letters},
	volume={19},
	number={11},
	pages={2005--2008},
	year={2015},
	publisher={IEEE}
}

@inproceedings{peralta2022outer,
	title={{Outer loop link adaptation enhancements for ultra reliable low latency communications in 5G}},
	author={Peralta, Elena and Pocovi, Guillermo and Kuru, Lauri and Jayasinghe, Keeth and Valkama, Mikko},
	booktitle={2022 IEEE 95th Vehicular Technology Conference:(VTC2022-Spring)},
	pages={1--7},
	year={2022},
	organization={IEEE}
}

@inproceedings{paymard2022enhanced,
	title={{Enhanced link adaptation for extended reality code block group based HARQ transmissions}},
	author={Paymard, Pouria and Amiri, Abolfazl and Kolding, Troels E and Pedersen, Klaus I},
	booktitle={2022 IEEE Globecom Workshops (GC Wkshps)},
	pages={711--716},
	year={2022},
	organization={IEEE}
}

@article{pulliyakode2017reinforcement,
	title={{Reinforcement learning techniques for outer loop link adaptation in 4G/5G systems}},
	author={Pulliyakode, Saishankar Katri and Kalyani, Sheetal},
	journal={arXiv preprint arXiv:1708.00994},
	year={2017}
}

@misc{ourcode,
	author       = {Maggi, Lorenzo and Wiesmayr, Reinhard and Cammerer, Sebastian and  Aoudia, Fay{\c{c}}al A{\"\i}t and Hoydis, Jakob and Keller, Alexander},
	title        = {{SALAD: Self-Adaptive Link Adaptation---Code Repository}},
	year         = {2025},
	howpublished         = {\url{https://github.com/NVlabs/salad}}
}

@article{kaltenberger2025driving,
	title={{Driving innovation in 6G wireless technologies: The OpenAirInterface approach}},
	author={Kaltenberger, Florian and Melodia, Tommaso and Ghauri, Irfan and Polese, Michele and Knopp, Raymond and Nguyen, Tien Thinh and Velumani, Sakthivel and Villa, Davide and Bonati, Leonardo and Schmidt, Robert and others},
	journal={Computer Networks},
	pages={111410},
	year={2025},
	publisher={Elsevier}
}

@article{nelder1965simplex,
	title={A simplex method for function minimization},
	author={Nelder, John A and Mead, Roger},
	journal={The computer journal},
	volume={7},
	number={4},
	pages={308--313},
	year={1965},
	publisher={The British Computer Society}
}

@article{cammerer2025sionna,
	title={{Sionna Research Kit: A GPU-Accelerated Research Platform for AI-RAN}},
	author={Cammerer, Sebastian and Marcus, Guillermo and Zirr, Tobias and Aoudia, Fay{\c{c}}al A{\"\i}t and Maggi, Lorenzo and Hoydis, Jakob and Keller, Alexander},
	journal={arXiv preprint arXiv:2505.15848},
	year={2025}
}

@book{fletcher2000practical,
	title={Practical methods of optimization},
	author={Fletcher, Roger},
	year={2000},
	publisher={John Wiley \& Sons}
}

@article{huang2021deluxe,
	title={{DELUXE: A DL-based link adaptation for URLLC/eMBB multiplexing in 5G NR}},
	author={Huang, Yan and Hou, Y Thomas and Lou, Wenjing},
	journal={IEEE Journal on Selected Areas in Communications},
	volume={40},
	number={1},
	pages={143--162},
	year={2021},
	publisher={IEEE}
}

@article{balakrishnan2002comparison,
	title={{A comparison of mechanisms for improving TCP performance over wireless links}},
	author={Balakrishnan, Hari and Padmanabhan, Venkata N and Seshan, Srinivasan and Katz, Randy H},
	journal={IEEE/ACM Transactions on Networking},
	volume={5},
	number={6},
	pages={756--769},
	year={2002},
	publisher={IEEE}
}

@article{carreras2018link,
	title={{Link abstraction models for multicarrier systems: A logistic regression approach}},
	author={Carreras Mesa, Alberto and Aguayo-Torres, Mari Carmen and Martin-Vega, Francisco J and Gomez, Gerardo and Blanquez-Casado, Francisco and -Luque, Isabel M and Entrambasaguas, Jose},
	journal={International Journal of Communication Systems},
	volume={31},
	number={1},
	pages={e3436},
	year={2018},
	publisher={Wiley Online Library}
}

@misc{nvidiaAerial,
	author       = {{NVIDIA Corporation}},
	title        = {{Aerial RAN CoLab Over-the-Air}},
	howpublished = {\url{https://docs.nvidia.com/aerial/aerial-ran-colab-ota/current/index.html}},
	note         = {Accessed: 2025-09-04}
}

@misc{cuphy,
	author       = {{NVIDIA Corporation}},
	title = {{cuPHY}},
	howpublished = {\url{https://docs.nvidia.com/aerial/aerial-cuphy/current/text/cuphy.html}},
	note         = {Accessed: 2025-09-04}
}

@article{hazan2016introduction,
	title={Introduction to online convex optimization},
	author={Hazan, Elad},
	journal={Foundations and Trends{\textregistered} in Optimization},
	volume={2},
	number={3-4},
	pages={157--325},
	year={2016},
	publisher={Now Publishers, Inc.}
}
	
\end{document}